%% file: FinalOneColArXiv.tex
\documentclass[12pt, draftclsnofoot, onecolumn]{IEEEtran}
\usepackage[T1]{fontenc}
\usepackage[pdftex]{graphicx}
\usepackage{epstopdf}
\usepackage{transparent}
\usepackage{eso-pic}

\input{header}

\newcommand{\Ac}{{\cal A}}
\newcommand{\Bc}{{\cal B}}
\newcommand{\Cc}{{\cal C}}
\newcommand{\Dc}{{\cal D}}

\newcommand{\Jc}{{\cal J}}

\newcommand{\Mc}{{\cal M}}

\newcommand{\Sc}{{\cal S}}

\newcommand{\Uc}{{\cal U}}

\newcommand{\Vc}{{\cal V}}

\usepackage{tikz}

\usepackage[utf8]{inputenc}
\usepackage{pgfplots} 
\usepackage{pgfgantt}
\usepackage{pdflscape}

\pgfplotsset{compat=newest} 
\pgfplotsset{plot coordinates/math parser=false}
\usepackage{relsize}
\usepackage{color}
\usepackage{changes}
\usepackage{graphicx,color,epsfig,rotating,subfigure}
\usepackage[cmintegrals]{newtxmath}
\makeatletter
\renewcommand*{\@opargbegintheorem}[3]{\trivlist
	\item[\hskip \labelsep{\bfseries #1\ #2}] \textbf{(#3)}\ \itshape}
\makeatother

\begin{document}
	
	\title{On the Optimality of D2D Coded Caching with Uncoded Cache Placement and One-shot Delivery}
	
	\author{
		\c{C}a\u{g}kan~Yapar,~\IEEEmembership{Student~Member,~IEEE,} 
		Kai~Wan,~\IEEEmembership{Member,~IEEE,} \\
		Rafael F. Schaefer,~\IEEEmembership{Senior~Member,~IEEE,}
		and~Giuseppe Caire,~\IEEEmembership{Fellow,~IEEE}
		
		\thanks{A short version of this paper was presented in the IEEE International Symposium on Information Theory (ISIT 2019) in Paris, France.
		}
		
		\thanks{
			\c{C}a\u{g}kan~Yapar, Kai Wan and Giuseppe Caire are with the Communications and Information Theory Chair, Technische Universit\"at Berlin, Germany
			(e-mail: cagkan.yapar@tu-berlin.de; kai.wan@tu-berlin.de; caire@tu-berlin.de). The work of \c{C}. Yapar, K.~Wan and G.~Caire was funded by the European Research Council  under the ERC Advanced Grant N. 789190, CARENET.}
		
		\thanks{Rafael F. Schaefer was with the Information Theory and Applications Chair, Technische Universit\"at Berlin, Germany. Currently, he is with the Chair of Communications Engineering and Security, University of Siegen, Germany (email: rafael.schaefer@uni-siegen.de). The work of R. F. Schaefer was funded by the German Ministry of Education and Research (BMBF) within the national initiative for “Post Shannon Communication (NewCom)” under Grant 16KIS1004.
		}

	}

	\IEEEoverridecommandlockouts
	\maketitle

	\begin{abstract}
		We consider a cache-aided wireless device-to-device (D2D) network of the type introduced by Ji, Caire, and Molisch \cite{ji2016fundamental},
		where the placement phase is orchestrated by a central server. 
		We assume that the devices' caches are filled with uncoded data, and the whole content database is contained in the collection of caches. 
		After the cache placement phase, the files requested by the users are serviced by inter-device multicast communication. 
		For such a system setting, we provide the exact characterization of the optimal load-memory trade-off under the assumptions of {\em uncoded placement}
		and {\em one-shot delivery}. In particular, we derive  both the minimum average (under uniformly distributed demands) 
		and the minimum worst-case sum-load of the D2D transmissions, 
		for given individual cache memory size at disposal of each user. 
		Furthermore, we show that the performance of the proposed scheme is within factor $4$ of the information-theoretic optimum.
		Capitalizing on the one-shot delivery property, we also propose an extension of the presented scheme that provides robustness against random user inactivity.
	\end{abstract}
	
	\section{Introduction}
	The killer application of wireless networks has evolved from real-time voice communication to on-demand multimedia content delivery (e.g., video), which requires a nearly $100$-fold increase in the per-user throughput, from tens of kb/s to  $1$ Mb/s.  Luckily, the pre-availability of such content allows for leveraging storage opportunities at users in a proactive manner, thereby reducing the amount of necessary data transmission during periods of high network utilization.
	
	A widely adopted  information theoretic model for a caching system (e.g., see \cite{maddah2014fundamental,ji2016fundamental,shariatpanahi2016multi,pedarsani2016online}) comprises two phases. 
	The \emph{placement phase} refers to the operation during low network utilization, when users are not requesting any content. During this phase, the cache memories of users are filled by a central server proactively. When each user directly stores a subset of bits, the placement phase is called \emph{uncoded}. The placement phase is called \emph{centralized} if the server knows the identity of the users in the system and coordinates the placement of the content based on this information. Otherwise, the placement without  such a coordination among the users is called \emph{decentralized}. 
	
	The transmission stage when users request their desired content is termed \emph{delivery phase}. By utilizing the content stored in their caches during the placement phase, the users aim to reconstruct their requested content from the signals they receive. The sources of such signals may differ depending on the context and network topology. In this work, we focus on the device-to-device (D2D) caching scenario, in which the signals available during the delivery phase are generated merely by the users themselves, whereas the central server remains inactive during the delivery phase. 
	
	A coded caching strategy was proposed  by Maddah-Ali and Niesen (MAN) \cite{maddah2014fundamental} for a prototypical network topology dubbed
	{\em single shared-link network}. Their model consists of users with caches and of a server which is in charge of the distribution of content to users through 
	an error-free shared link, such that the server transmission is multicasted to all the users. 
	The system performance is characterized in terms of the {\em load} of the shared link (i.e., the length, in coded bits, of the multicast message necessary such that
	all the users can successfully decode their requested content item) as a function of the number of users, cache memory size per user, and  size of the 
	content library.   This seminal work showed that a \emph{global caching gain} is possible by utilizing  multicast linear combinations during the delivery phase, whereas the previous work on caching \cite{baev2008approximation,almeroth1996use,dan1996dynamic,korupolu2001placement,meyerson2001web,dowdy1982comparative} aimed to benefit from the \emph{local caching gain}, omitting the multicasting opportunity. 
	
	By observing that some MAN linear combinations are redundant,
	an improved delivery scheme was proposed in \cite{yu2018exact}, which was proved to be information theoretically optimal 
	(i.e., it achieves an information theoretic converse bound on the minimum possible load of the shared link) 
	under the constraint of uncoded cache placement with respect to two performance criteria:  a) the average load under random uniform demands; b) the worst-case load with respect to any possible demand configuration. 
	Also, it was proved in~\cite{yu2017characterizing} that such scheme 
	is optimal within a factor of $2$ over all possible schemes, even removing the constraint of uncoded placement.
	
	The work \cite{maddah2014fundamental} has generated a lot of attention in the information theoretic and network-coding theoretic 
	community and led to numerous extensions, e.g., decentralized caching \cite{maddah2015decentralized}, device-to-device (D2D) caching \cite{ji2013fundamental,ji2016fundamental,ibrahim2018device}, caching on file selection networks \cite{lim2017information}, caching with nonuniform demands
	\cite{niesen2017coded,ji2017order,zhang2015coded}, multi-server \cite{shariatpanahi2016multi}, online caching \cite{pedarsani2016online}, just 
	to name some.
	
	The D2D caching problem was originally considered in \cite{ji2013fundamental,ji2016fundamental,ibrahim2018device}, where  users are allowed to  communicate with each other. By extending  the caching scheme in \cite{maddah2014fundamental} to the D2D scenario, a global caching gain can also be achieved. It was proved in \cite{ji2013fundamental} and \cite{ji2016fundamental} that the proposed D2D caching scheme is order optimal within a constant factor
	when the memory size is not small.
	
	Particularly, the D2D caching setting with uncoded placement considered in this work is closely related to the problem of distributed computing   \cite{li2018fundamental,ji2018fundamental,li2018wireless,li2017scalable,reisizadeh2017coded,yu2017optimally,kiamari2017heterogeneous,bitar2017minimizing} and data-shuffling problems  \cite{wan2018fundamental,song2017pliable}. The coded distributed computing setting can be interpreted 
	as a symmetric D2D caching setting with multiple requests, whereas the coded data shuffling problem can be viewed as a D2D caching 
	problem with additional constraints on the placement.
	
	So far, the exact information theoretic optimality for a D2D caching scheme (possibly under some constraints), parallel to the 
	results  of \cite{yu2018exact} for the shared-link network are open. This paper partially closes this problem by exhibiting an explicit D2D centralized
	coded caching scheme with exactly optimal performance (i.e., matching an information theoretic converse bound) under the constraint
	of uncoded placement and one-shot delivery.\footnote{See Definition \ref{one-shot} for a formal definition of one-shot delivery 
		and its relevance in the context of D2D networks.}

	\subsection{Our Contributions}
	The main contributions in this paper are:
	\begin{enumerate}
		\item
		Based on the D2D achievable caching scheme in~\cite{ji2016fundamental}, and the improved delivery scheme 
		for the shared-link network in \cite{yu2018exact}, we propose a novel achievable scheme for D2D caching problem, which is shown to be order optimal within a factor of $2$ under the constraint of uncoded placement and within a factor of $4$ in general, in terms of the average load with random uniform demands and 
		worst-case load over all possible demands, while in the literature there is no exactly optimality on the D2D coded caching problem and the lowest order optimality factor is $8$ proved in \cite{sengupta2015beyond}.
		\item 
		We say that the delivery phase is ``one-shot'' if any user can recover any requested bit from the content if its own cache and at most one transmitted message by some other user (see Definition \ref{one-shot}).   
		Under the constraint of uncoded placement and one-shot delivery, we can divide the D2D caching problem into $K$ shared-link models. Under the above constraints, we then use the index coding acyclic converse bound in~\cite[Corollary 1]{onthecapacityindex}  to lower bound the total load transmitted in the $K$ shared-link models. By leveraging the connection among the $K$ shared-link models, we propose a novel way to use the index coding acyclic converse bound compared to the method used for single shared-link model in~\cite{wan2016optimality,wan2016caching,yu2018exact}. With this converse bound, we prove that the proposed D2D caching scheme is {\em exactly optimal} under the constraint of uncoded placement and one-shot delivery, both in terms of the average 
		load under random uniform demands and in terms of the worst-case load over all possible demands.
		\item
		Inspired by the distributed computing problem with \emph{straggler}s (see e.g. \cite{speedingUpML} for a distributed linear computation scenario), 
		where straggling servers fail to finish their computational tasks on time,
		we design a coded caching scheme for a D2D scenario where, during the delivery phase, each user may be inactive with some probability 
		and the inactivity event of each user is not known by other users. User inactivity may occur due to several reasons such as broken communication links, users moving out of the network, users going off-line to save power, etc. For this setting, a non-one-shot delivery scheme would be very fragile since, because of the fact that requested bits are decoded from the transmissions of multiple users, a missing user may affect the decoding of many packets, 
		through a sort of catastrophic error propagation effect.
		Instead, we can directly extend the proposed optimal one-shot delivery phase to this problem by using {\em Maximum Distance Separable} (MDS) 
		precoding proposed in \cite{speedingUpML}, which promotes robustness against random unidentified user inactivity.
		
	\end{enumerate}

	The rest of this paper is organized as follows. We provide a precise definition of our model and an overview of previous results on D2D and shared-link caching scenarios in Section \ref{sec:sys}. Section \ref{sec:RMtrade-off} presents our main new results. 
	The proposed caching scheme is detailed in Section \ref{sec:achiev}. 
	The proof of optimality under the constraint of one-shot delivery is obtained via a matching converse in Section \ref{sec:Converse}. 
	We treat the problem of random user inactivity by extending the presented scheme in Section \ref{sec:loadouta}. 
	Finally, we illustrate our results with the aid of numerical comparisons with the existing bounds in Section \ref{sec:Numerical}.
	
	\section{Problem Setting and Related Background}\label{sec:sys}

	In this section, we define our notations and network model and present previous results which are closely related to the problem we consider in the current work. 
	
	\subsection{Notation}
	\label{sec:model:notation}
	$|\cdot|$ is used to represent the cardinality of a set or the length of a file in bits;
	we let 
	$\mathcal{A\setminus B}:=\left\{ x\in\Ac|x\notin\mathcal{B}\right\}$,
	$[a:b:c]:=\{a,a+b,a+2b,...,c\}$,
	$[a:c] = [a:1:c]$ and $[n]=[1:n]$;
	the bit-wise XOR operation between binary vectors is indicated by $\oplus$;
	for two integers $x$ and $y$, if $x<y$ or $x\leq0$, we let $\binom{x}{y}=0$; $\mathbb{1\{\cdot\}}$ denotes the indicator function.

	\subsection{D2D Caching Problem Setting}
	\label{sub:problem setting}
	\begin{figure}
		\centering{}
		\scalebox{0.85}{\includegraphics[scale=1,trim=0.1cm 0cm 0.1cm 0cm,clip]{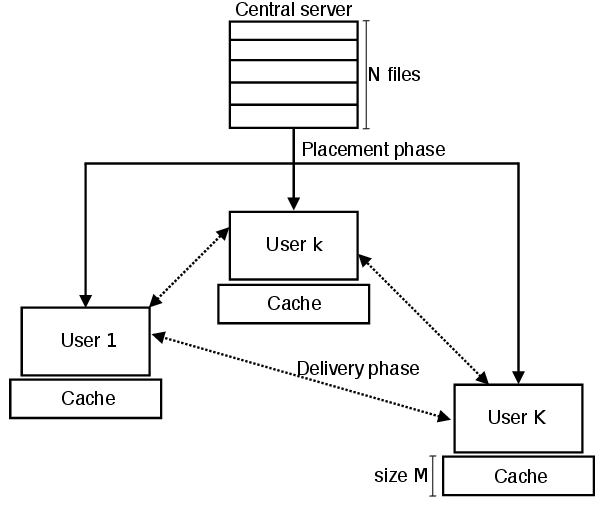}}
		\caption{System model for cache-aided D2D network where, during the delivery phase, the users produce coded messages from the bits stored in their own caches and broadcast such messages to all other users via a shared medium (e.g., a wireless channel).
			At the end of the delivery phase, each user must be able to decode its own requested content item from its own cache and the collection of all messages
			received from the other users.  Solid and dotted lines indicate operation during placement and delivery phases, respectively.}
		\label{fig:scheme}
	\end{figure}
	We consider a D$2$D network composed of $K$ users, which are able to receive all the other users' transmissions (see Fig. \ref{fig:scheme}). We assume a collision avoidance protocol for which when a user transmits, all the other stay quiet and listen (e.g., this can be implemented in a practical wireless network using CSMA, as in 
	the IEEE 802.11 standard). 
	The system comprises  also a server with a fixed content file database of $N$ files $\boldsymbol{\mathcal{W}} := (W_1,\dots, W_N)$ each with a length of $F$ bits. 
	Every user has a cache memory of $MF$ bits, $M < N$, at its disposal. Following the classical information theoretic coded caching paradigm
	\cite{maddah2014fundamental,ji2016fundamental}, the system operation can be divided into two phases, namely, 
	the \emph{placement} and \emph{delivery} phases.
	
	During the placement phase the central server places data produced from the content library $\boldsymbol{\mathcal{W}}$ into the user caches.\footnote{A more general
		coded placement would consider cache encoding functions $\phi_k : \boldsymbol{\mathcal{W}} \mapsto Z_k$, producing the cache content $Z_k$ of user $k$ subject to the entropy constraint $H(Z_k) \leq MF$.} 
	In this work, we consider {\em uncoded placement}  \cite{yu2018exact},  where each user $k$ directly stores $MF$ bits from the $NF$ total bits $\boldsymbol{\mathcal{W}}$. 
	This assumption is implicit in the rest of the paper and will not be repeated for the sake of brevity. 
	Since the placement is uncoded, we can divide each file $W_q$, $q \in [N]$, into sub-files $W_{q} = \{W_{q,\Vc}:\Vc\subseteq [K]\}$, 
	where $W_{q,\Vc}$ represents the set of bits of file $q$ {\em exclusively} cached by users in $\Vc \subseteq [K]$.

	We denote the collection of the stored bits at user $k$ by $\mathcal{M}_k$ and the cache placement of the whole system by $\boldsymbol{\mathcal{M}} := (\mathcal{M}_1,\dots,\mathcal{M}_K)$. We assume that, at the end of this phase, each bit of the database is available at least 
	in one of the users' caches, i.e., implying $\boldsymbol{\mathcal{W}} \subseteq \boldsymbol{\mathcal{M}}$, otherwise, there would be 
	files that cannot be retrieved using only D2D transmissions. 
	This obviously implies that $MK \geq N$ must hold.
	
	During the delivery phase, each user requests one file. We define the {\em request vector} $\boldsymbol{d}:=\left(d_1,\dots,d_K\right)$, with $d_k \in [N]$ denoting user $k$'s requested file index. The set of all possible request vectors is denoted by $\mathcal{D}$, so that $\mathcal{D}=[N]^K$.
	Following the classical information theoretic coded caching paradigm, 
	we assume that the requests are revealed to all users via some control channel (e.g., coordinated by the server). 
	Since the amount of bits necessary to notify the requests is much less than the actual requested data delivery, the overhead incurred by the request broadcasting 
	can be neglected \cite{maddah2014fundamental,ji2016fundamental,shariatpanahi2016multi,pedarsani2016online,yu2018exact,yu2017characterizing,niesen2017coded,ji2017order,zhang2015coded}. 
	Given the request vector, each user $k$ generates a codeword $X_k$ of length $R_k F$ bits and broadcasts it to other users, where $R_k$ indicates the load of user $k$. For a given subset of users $\mathcal{T} \subseteq [K]$, we let $X_{\mathcal{T}}$ denote the ensemble of codewords broadcasted by these users. From the stored bits in $\mathcal{M}_k$ and the received codewords $X_{[K]\backslash k}$, each user $k$ attempts to recover its desired file $W_{d_k}$.
	More specifically, at the end of the delivery phase each user $k$ needs to decode all the bits of file $W_{d_k}$ not already 
	contained in its own cache $\mathcal{M}_k$. These bits, referred to as the {\em needed bits} of user $k$, 
	must be retrieved from  $\mathcal{M}_k$ and the ensemble of transmitted codewords $X_{[K]\backslash k}$.
	
	In this work we concentrate on the special case of \emph{one-shot delivery}, i.e., 
	such that any user can recover any requested bit from the content if its own cache and at most one transmitted message by some other user.
	The formal mathematical  definition is given in the following.
	\begin{definition}[One-shot delivery] \label{one-shot}
		Let $W^{k}_{d_k}(i)$ denote the $i$-th needed bit of user $k$. If for each $i$ there exists some user index $j_k(i)$ such that
		\[ H(W^{k}_{d_k}(i) | X_{j_k(i)}, \Mc_k) = 0, \]
		then the delivery scheme is called ``one-shot''. 	In words, this means that 
		$W^{k}_{d_k}(i)$ is a deterministic function of the codeword transmitted by a {\em single} user $j_k(i)$ and of the cache content $\Mc_k$ of the demanding user $k$. Therefore, each user $k$ can recover each of its needed bits from its own cache and the transmission of a single other user. 
		\hfill $\lozenge$
	\end{definition}
	
	Under the one-shot delivery assumption, let $W^{k,k'}_{d_k}$ denote the block of all bits needed by user $k$ and decoded from the codeword transmitted by a user $k'$.
	Then, the one-shot condition implies
	\[ (W_{d_k} \setminus \Mc_k) \subseteq \bigcup_{k' \in [K]\setminus \{k\}} W^{k,k'}_{d_k}. \]
	In addition, we also define $W^{k,k'}_{d_k,\Vc}$ as the block of bits needed by user $k$ and recovered from the codeword 
	of user $k'$ which are exclusively cached by users in $\Vc$. Hence, for all pairs of users $k,k' \in[K]$, the following holds
	\[ \bigcup_{\Vc\subseteq ([K]\setminus \{k\}): k' \in \Vc} W^{k,k'}_{d_k,\Vc} = W^{k,k'}_{d_k}.\]
	
	\begin{remark}\label{rem:one-shot background}
		\emph{One-shot} terminology is often used in settings related to interference alignment in Gaussian interference channels. 
		To the best of our knowledge, the only work which explicitly emphasized the one-shot delivery (same as in Definition \ref{one-shot})
		in the caching  setting before the present work was \cite{naderializadeh2017fundamental}. 
		Notice that one-shot does not mean that each user recovers {\em all} its needed bits from the transmission of a single other user. 
		For example, different needed bits $W^{k}_{d_k}(i)$ and $W^{k}_{d_k}(i')$ with $i \neq i'$ may be recovered from different codewords
		$X_{j_k(i)}$ and $X_{j_k(i')}$ with $j_k(i) \neq j_k(i')$. Nevertheless, as soon as user $j_k(i)$ broadcasts its codeword, user $k$
		can decode its needed bit $W^{k}_{d_k}(i)$ immediately, without waiting for the transmission of other users. 
		\hfill $\lozenge$
	\end{remark}
	
	\begin{remark}\label{rem:one-shot motivation}
		One-shot delivery is indeed very desirable in a D2D setting. In fact, in a non-one-shot scheme, some needed bits can only be decoded from 
		the transmissions of multiple users (otherwise the scheme would be one-shot). 
		This requires to store such multiple transmissions at the users' receiver, and wait for all these other users to complete their broadcasts
		before such bits can be retrieved. For example, in Section \ref{sec:loadouta}  we shall design a coded scheme that is able to cope with 
		random user inactivity. This is  relatively immediate in the case of one-shot delivery, while it would be much more complicated for non-one-shot delivery schemes. Furthermore, 
		we show in Theorem \ref{thm:order optimality} of this paper that any scheme based on non-one-shot delivery and/or general coded placement 
		can improve the system performance by at most  a factor of 2. Therefore, investigating non-one-shot delivery schemes and/or going away from 
		the well-understood uncoded placement has arguably a limited practical interest in terms of possible performance improvement.
		\hfill $\lozenge$
	\end{remark}

	Letting $R = \sum_{k=1}^{K} R_k$, we say that a communication load $R$ is \textit{achievable} for a request vector $\boldsymbol{d}$ and placement $\boldsymbol{\mathcal{M}}$, with $\abs{\mathcal{M}_k}=M,\, \forall k \in [K]$, if there exists an ensemble of codewords $X_{[K]}$ of size $RF$ such that each user $k$ can reconstruct its requested file $W_{d_k}$. We let $R^*(\boldsymbol{d},\boldsymbol{\mathcal{M}})$ indicate the minimum achievable load given $\boldsymbol{d}$ and $\boldsymbol{\mathcal{M}}$.
	We also define $R^*_{\textup{o}}(\boldsymbol{d},\boldsymbol{\mathcal{M}})$ as the minimum  achievable load given $\boldsymbol{d}$ and $\boldsymbol{\mathcal{M}}$ under the constraint of one-shot delivery.
	
	We consider random user demands, i.e., $\boldsymbol{d}$ is distributed on $\mathcal{D}$ according to some probability distribution.
	Given a placement $\boldsymbol{\mathcal{M}}$, the average load $R^*_{\textup{ave}}(\boldsymbol{\mathcal{M}})$ is defined 
	as the expected  minimum achievable load under the given distribution of requests:	
	\begin{equation*}
		R^*_{\textup{ave}}(\boldsymbol{\mathcal{M}})=\mathbb{E}_{\boldsymbol{d}}[ R^*(\boldsymbol{d},\boldsymbol{\mathcal{M}})].
	\end{equation*}
	We define $R^*_{\textup{ave}}$ as the minimum achievable average load:
	\begin{equation*}
		R^*_{\textup{ave}}=\min_{\substack{\boldsymbol{\mathcal{M}}}} R^*_{\textup{ave}}(\boldsymbol{\mathcal{M}}).
	\end{equation*}
	Similarly, we define $R^*_{\textup{ave, o}}$ as the minimum average load under the constraint of one-shot delivery.
	
	Furthermore, for a given placement $\boldsymbol{\mathcal{M}}$, the worst-case load $R_{\textup{worst}}^*(\boldsymbol{\mathcal{M}})$ is defined as 
	\begin{equation*}
		R_{\textup{worst}}^*(\boldsymbol{\mathcal{M}})=\max_{\boldsymbol{d}} R^*(\boldsymbol{d},\boldsymbol{\mathcal{M}}).
	\end{equation*}	
	We define $R^*_{\textup{worst}}$ as the minimum achievable worst-case load:
	\begin{equation*}
		R_{\textup{worst}}^*=\min_{\boldsymbol{\mathcal{M}}} R_{\textup{worst}}^*(\boldsymbol{\mathcal{M}}).
	\end{equation*}
	Correspondingly, we define $R^*_{\textup{worst, o}}$ as the minimum average load under the constraint 
	of one-shot delivery.
	
	Following the notation of \cite{yu2018exact}, we let $N_{\textup{e}}(\boldsymbol{d})$ denote the number of distinct elements (i.e., distinct files) 
	in a request vector  $\boldsymbol{d}$. In addition,
	we let $\boldsymbol{d}_{\backslash\{k\}}$ and $N_{\textup{e}}(\boldsymbol{d}_{\backslash\{k\}})$ stand for the request vector of users $[K]\backslash\{k\}$ and the number of distinct files requested by all users but user $k$, respectively. 
	
	\subsection{Main Background Results}\label{sec:Related}
	In this subsection, we briefly summarize the related approaches, namely the optimal shared-link scheme in \cite{yu2018exact} and the approach in the fundamental D2D work \cite{ji2016fundamental}, which are essential for appreciating our results for the D2D model.
	To this end, we first introduce the uncoded symmetric placement scheme which was introduced in the seminal work \cite{maddah2014fundamental}.

	\begin{definition}[{Maddah-Ali Niesen $\text{[MAN]}$ Symmetric Uncoded Placement Scheme}] \label{def:MANplacement}
		When $M=tN/K$, $t \in [0:K]$, each file $W_q$ is divided into $\binom{K}{t}$ disjoint sub-files denoted by $W_{q, \Vc}$, where $\Vc \subseteq [K]$ and $|\Vc| = t$. Each user $k$ caches all the bits of sub-files $W_{q, \Vc}$, $q \in [N]$, for all  $\Vc \ni k$. 
		Since there are $\binom{K-1}{t-1}$ sub-files for each file with $\Vc \ni k$ and each sub-file is composed of $F/\binom{K}{t}$ bits, 
		each user caches $NFt/K = MF$ bits, which fulfills the memory constraint. \hfill $\lozenge$
	\end{definition}
	
	\subsubsection{An Optimal Scheme for the Shared-link Coded Caching Problem}\label{sub:sharedLinkRelated}
	In the shared-link model \cite{maddah2014fundamental}, (also referred to as the \emph{bottleneck} model), a server with $N$ files is connected to $K$ users through an error-free channel. Each file is composed of $F$ bits and each user is provided with a local cache of size $MF$ bits. 
	For uncoded placement, the minimum average and worst-case loads are given as follows \cite{yu2018exact}:
	\begin{theorem}\label{teoSL}
		For the single shared-link network with a database of $N$ files, $K$ users each with a cache of size $M$, 
		and random uniformly distributed demands\footnote{This means that $\boldsymbol{d}$ is a random vector uniformly distirbuted over $\mathcal{D}$.
			Interestingly, exact optimality for the average load in the case of random non-uniformly distributed demands is still an open problem
			even for the single shared-link network.}
		the following average load $R^{\textup{sl*}}_{\textup{ave}}$  is optimal under the constraint of uncoded cache placement:
		\begin{equation}
		R^{\textup{sl*}}_{\textup{ave}}=\mathbb{E}_{\boldsymbol{d}}\left[ \frac{\binom{K}{t+1}-\binom{K-N_{\textup{e}}(\boldsymbol{d})}{t+1}}{\binom{K}{t}}\right],\label{eq:averageworstcaseSL}
		\end{equation}
		with $t=\frac{KM}{N}\in [K]$. 
		When $t\notin [K]$, $R^{\textup{sl*}}_{\textup{ave}}$ is given by the lower convex envelope of the values in 
		(\ref{eq:averageworstcaseSL}) for integer values of $t\in [K]$.  \hfill $\square$
	\end{theorem}
	
	\begin{corollary}
		\label{corrSL}
		For the same network as in Theorem \ref{teoSL} the following worst-case load $R^{\textup{sl*}}_{\textup{worst}}$ is optimal 
		under the constraint of uncoded cache placement:
		\begin{align}
			R^{\textup{sl*}}_{\textup{worst}}=  \frac{\binom{K}{t+1}-\binom{K-\min\{K,N\}}{t+1}}{\binom{K}{t}},
			\label{eq:worstCaseloadSL}
		\end{align}
		with $t=\frac{KM}{N}\in [K]$. When $t\notin [K]$, $R^{\textup{sl*}}_{\textup{worst}}$ is given by the lower convex envelope of the values in 
		(\ref{eq:worstCaseloadSL}) for integer values of $t\in [K]$.  \hfill $\square$
	\end{corollary}
	
	Notice that for $K \leq N$, the worst-case request has all distinct files and the negative term in (\ref{eq:worstCaseloadSL}). 
	The achievability for this case was in fact already presented in the seminal paper by Maddah-Ali and Niesen \cite{maddah2014fundamental} and its optimality was first proved in \cite{wan2016optimality}. In the general case, the optimal achievable scheme is based on the MAN symmetric uncoded placement. 
	However, in delivery phase, any subset of users that request $N_{\textup{e}}(\boldsymbol{d})$ distinct files is 
	identified as $\mathcal{U}=\{u_1,...,u_{N_{\textup{e}}(\boldsymbol{d})}\}$. These users are referred to 
	as \textit{leaders} \cite{yu2018exact}. The server broadcasts the codeword 
	\begin{equation}\label{eq:BroadcastsSL}
	Y_{\mathcal{A}}:=\underset{k\in \mathcal{A}}{{\bigoplus}}W_{d_k, \mathcal{A} \backslash\{k\}},
	\end{equation} 
	for all the subsets $\mathcal{A}\subseteq [K]$ of size $t+1$ such that $\mathcal{A}\cap\mathcal{U}\neq\emptyset$. As shown in \cite{yu2018exact}, the codewords $Y_{\mathcal{A'}}$, $\mathcal{A'}\cap\mathcal{U} = \emptyset$, that are not broadcasted by the server, can be recovered from the transmitted codewords.
	Thus, any user $k$ can decode any sub-file $W_{d_k, \mathcal{B}}$ it requires, where $\Bc \subseteq [K]\backslash\{k\}$, $|\Bc| = t$, from $Y_{\Bc \cup \{k\}}$, which is either broadcasted or reconstructed from the broadcasted codewords, by performing 
	\begin{equation}\label{eq:SLBroadcast}
	W_{d_k, \mathcal{B} } =  \left(\underset{x\in \mathcal{B}}{{\bigoplus}}W_{d_x, \{\mathcal{B} \cup \{k\} \}\backslash\{x\}}\right) \bigoplus Y_{\Bc\cup \{k\}} 
	\end{equation}
	as user $k$ has all the sub-files $W_{d_x, \{\mathcal{B} \cup \{k\} \}\backslash\{x\}}$, $\forall x\in \mathcal{B}$ and $Y_{\Bc\cup \{k\}} = \left(\underset{x\in \mathcal{B}}{{\bigoplus}}W_{d_x, \{\mathcal{B} \cup \{k\} \}\backslash\{x\}}\right) \bigoplus W_{d_k, \mathcal{B} }$.

	Overall, the server broadcasts $\binom{K}{t+1}-\binom{K-N_{\textup{e}(\boldsymbol{d})}}{t+1}$ codewords, so that the achieved load by this scheme for the request vector $\boldsymbol{d}$ amounts to
	\begin{align}\label{eq:singleSL}
		R^{\textup{sl*}}(\boldsymbol{d},\boldsymbol{\boldsymbol{\mathcal{M}}_\textup{MAN}})=\frac{\binom{K}{t+1}-\binom{K-N_{\textup{e}}(\boldsymbol{d})}{t+1}}{\binom{K}{t}},
	\end{align}
	where $\boldsymbol{\mathcal{M}}_\textup{MAN}$ refers to the symmetric uncoded placement.
	For non-integer $t$ with $1 < t < K$, the lower convex envelope of these points can be achieved by memory sharing.

	Applying this caching scheme for each request vector $\boldsymbol{d}$, the presented minimum loads can be achieved.
	\subsubsection{Graphical Converse Bound for the Shared-link Coded Caching Problem}
	\label{sub:shared-link converse}
	We now briefly summarize how the matching lower bound for the shared-link model can be derived by exploiting a connection
	(first proposed in \cite{wan2016optimality,wan2016caching}) between the single shared-link caching problem with uncoded cache placement and {\em index coding}.
	For any uncoded cache placement and request vector $\mathbf{d}$, we can generate a directed graph containing 
	a node for each sub-file demanded by each user and a directed edge from node $i$ to node $j$ if and only if the user demanding 
	the sub-file represented by node $j$ caches the sub-file represented by node $i$. 
	As a direct consequence of  the acyclic index coding converse bound  \cite[Corollary 1]{onthecapacityindex},  
	if the subgraph over a set of nodes $\Jc$ does not contain a directed cycle, letting the set of sub-files corresponding to these nodes be denoted by $\Sc_{\Jc}$ and the length of each sub-file $i$ by $L(i)$, the broadcast load (denoted by $R^{\textup{sl}*}(\mathbf{d})$) is lower bounded by
	\begin{align}
		R^{\textup{sl}*}(\mathbf{d}) \geq \sum_{i\in \Sc_{\Jc}} L(i).\label{eq:general acyclic bound}
	\end{align}
	
	The authors in~\cite{wan2016optimality,wan2016caching} proposed a way to choose the maximal acyclic sets in the graph.  We choose $N_{\textup{e}}(\mathbf{d})$ users with different demands. The chosen user set
	is denoted by $\Cc=\{c_{1},c_{2},...,c_{N_{\textup{e}}(\mathbf{d})}\}$ where  $c_{i}\in[K]$. 
	Each time, we consider a permutation of $\Cc$, denoted by $\mathbf{u}=(u_{1},u_{2},...,u_{N_{\textup{e}}(\mathbf{d})})$. It was proved in~\cite[Lemma 1]{wan2016optimality} that the following set of sub-files is acyclic, $\big(W_{d_{u_{i}},\Vc_{i}} : 
	\Vc_{i}\subseteq[K]\backslash \{u_{1},\ldots,u_{i}\},
	\ i\in[N_{\textup{e}}(\mathbf{d})]
	\big)$.
	By using~\eqref{eq:general acyclic bound}, we have
	\begin{align}
		R^{\textup{sl}*}(\mathbf{d})
		&\geq \sum_{i\in[N_{\textup{e}}(\mathbf{d})]} \sum_{\Vc_{i}\subseteq[K]\backslash\{u_{1},\ldots,u_{i}\}}   \frac{ |W_{d_{u_{i}},\Vc_{i}}| }{F}.
		\label{eq:original uncycle}
	\end{align}
	
	Considering all the possible sets of the $N_{\textup{e}}(\mathbf{d})$ users with different demands and all the permutations, we sum all the inequalities in 
	the form of~\eqref{eq:original uncycle} to derive a converse bound of  $R^{\textup{sl}*}(\mathbf{d})$ in terms of the lengths of  sub-files. The next step is to consider all the possible demands and use the Fourier-Motzkin algorithm \cite{el2011network} to eliminate all the sub-files with the constraints of cache size and file size. Following this approach, we can derive a tight converse bound for the shared-link model.
	
	\subsubsection{The Previous Approach \cite{ji2016fundamental} for the D2D Coded Caching Problem }\label{sub:D2DRelated}
	Also in this case, the achievable scheme is based on the MAN symmetric uncoded placement. 
	Further, each sub-file is divided into $t$ equal length disjoint sub-pieces of $F/ t \binom{K}{t}$ bits which are denoted by $W_{q, \Vc, i}$, 
	where $i\in \Vc$. During the delivery phase, each user $i$ broadcasts the codeword 
	\begin{equation}\label{eq:BroadcastsSLD2D}
	Y_{\mathcal{A}^i}^i:=\underset{k\in \mathcal{A}^i}{{\bigoplus}}W_{d_k, \{\mathcal{A}^i \cup \{i\}\}\backslash\{k\}, i},
	\end{equation}
	for all the subsets $\mathcal{A}^i\subseteq [K]\backslash\{i\}$ of $t$ users.
	Hence, any user $k$ can decode any needed sub-piece $W_{d_k, \mathcal{B}^k \cup \{i\},i}$, 
	where $\Bc^k \subseteq [K]\backslash\{i,k\}$, $|\Bc^k| = t-1$, from $Y^i_{\Bc^k \cup \{k\}}$, which is broadcasted from user $i$, by performing
	\begin{equation}\label{eq:decodeBroadcastJCM}
	W_{d_k, \mathcal{B}^k \cup \{i\},i} =  \left(\underset{x\in \mathcal{B}^k}{{\bigoplus}}W_{d_x, \{\mathcal{B}^k \cup \{i,k\}\}\backslash\{x\}, i}\right) \bigoplus Y^i_{\Bc^k\cup \{k\}} 
	\end{equation}
	as user $k$ has all the sub-pieces $W_{d_x, \{\mathcal{B}^k \cup \{i,k\}\}\backslash\{x\}, i}$, $\forall x\in \mathcal{B}^k$  and $Y^i_{\Bc^k\cup \{k\}}  = \left(\underset{x\in \mathcal{B}^k}{{\bigoplus}}W_{d_x, \{\mathcal{B}^k \cup \{i,k\}\}\backslash\{x\}, i}\right) \bigoplus W_{d_k, \mathcal{B}^k \cup \{i\},i}$.
	Notice that the above delivery scheme is one-shot, following Definition \ref{one-shot}. 
	
	Each user broadcasts a total of $\binom{K-1}{t}$ codewords. hence,  the achieved load is given by
	\begin{align}
		R_{\textup{JCM}} = \frac{K\binom{K-1}{t}}{t \binom{K}{t}} =  \frac{N}{M}-1.
	\end{align}
	Moreover, for non-integer $t$ with $1 < t < K$, the lower convex envelope of these points is achievable by memory sharing.
	This load was shown \cite{sengupta2015beyond} to be optimal within a factor of $8$, even when coding in the placement phase is taken into consideration.
	
	
	
	\section{Main Results}\label{sec:RMtrade-off}
	
	In this section we state the main results of this work. In the following theorem, we characterize the exact memory-average load trade-off of the D2D caching network
	with random uniform demands and under the constraint of one-shot delivery. 
	The achievable scheme is presented in Section~\ref{sec:achiev} and the converse bound is proved in Appendix~\ref{sec:Converse}.
	
	\begin{theorem}[Average load]\label{teo}
		For the D2D caching network as defined in Section \ref{sub:problem setting}, with a database of $N$ files and $K$ users each with a cache of size $M$,  
		the following average load is optimal for random uniform demands and under the constraint of uncoded placement and one-shot delivery:
		\begin{equation}
		R^*_{\textup{ave, o}}=\mathbb{E}_{\boldsymbol{d}}\left[ \frac{\binom{K-1}{t}-\frac{1}{K}\sum_{k=1}^K\binom{K-1-N_{\textup{e}}(\boldsymbol{d}_{\backslash\{k\}})}{t}}{\binom{K-1}{t-1}}\right]\label{eq:averageworstcase}
		\end{equation}
		with $t=\frac{KM}{N} \in [K]$, where $\boldsymbol{d}$  is uniformly distributed over $\mathcal{D}=[N]^K$. 
		When $t\notin [K]$, $R^*_{\textup{ave, o}}$ is given by the lower convex envelope of the values in 
		(\ref{eq:averageworstcase}) for integer values of $t \in [K]$.  \hfill $\square$
	\end{theorem}
	For the worst-case load criterion, we have:
	\begin{corollary}[Worst-case load]
		\label{corr}
		For the D2D caching network as in Theorem \ref{teo}, the following worst-case load $R^*_{\textup{worst, o}}$ is optimal 
		under the constraint of uncoded placement and one-shot delivery:
		\begin{align}
			R^*_{\textup{worst, o}}	&= \max_{\boldsymbol{d}} \frac{\binom{K-1}{t}-\frac{1}{K}\sum_{k=1}^K\binom{K-1-N_{\textup{e}}(\boldsymbol{d}_{\backslash\{k\}})}{t}}{\binom{K-1}{t-1}}\label{eq:worseImplicit}\\
			&= \begin{cases} \frac{\binom{K-1}{t}}{\binom{K-1}{t-1}} & K \leq N  \\  
				\frac{\binom{K-1}{t}-\frac{2N-K}{K}\binom{K-N}{t}-\frac{2(K-N)}{K}\binom{K-1-N}{t}}{\binom{K-1}{t-1}} & \textup{otherwise}\\
				\frac{\binom{K-1}{t}-\binom{K-1-N}{t}}{\binom{K-1}{t-1}} & K \geq 2N
			\end{cases}\label{eq:worstCaseload}
		\end{align}
		with $t = \frac{KM}{N} \in [K]$. 
		When $t\notin [K]$, $R^*_{\textup{worst, o}}$ is given by the lower convex envelope of the values in 
		(\ref{eq:worstCaseload}) for integer values of $t \in [K]$.  \hfill $\square$	
	\end{corollary}
	
	\begin{IEEEproof}
		The load stated in \eqref{eq:worseImplicit} can be achieved by the scheme presented in Section~\ref{sec:achiev} and its optimality is proved 
		in  Appendix~\ref{sec:Converse}. The explicit characterization of the worst-case demand which gives \eqref{eq:worstCaseload} 
		can be found in Appendix~\ref{sec:worstDemand}. 
	\end{IEEEproof}
	%
	%
	
	By comparing the load achievable by our scheme and the minimum achievable load for the shared-link model, which is trivially a lower bound on the best possible load
	achievable in the D2D case under the system definition of Section \ref{sub:problem setting}, we obtain the following order optimality result, which will be proved in Appendix~\ref{sec:orderOpt}.
	
	\begin{theorem}[Order optimality]\label{thm:order optimality}
		For the D2D caching network as in Theorem \ref{teo}, the achievable average (under uniform demands) and worst-case loads in~\eqref{eq:averageworstcase} and~\eqref{eq:worstCaseload} 
		are order optimal within a factor of $2$  under the constraint of uncoded placement and within a factor of $4$ in general. 
	\end{theorem}

	\begin{remark}
		In \cite{ibrahim2018device}, a lower bound for the D2D caching problem with $K \leq N$ under the constraint of uncoded placement is developed by treating the D2D problem as set of K shared-link caching networks where, in turns, each user act as the server and satisfies a subset of the requested needed bits of the other users. Then, the index-coding lower bound can be used for each shared-link model, similar to the approach proposed in the current work. However, the problem can be decomposed into K shared-link caching models only under the assumption of one-shot delivery. In the absence of such constraint, the bounds proposed in \cite{ibrahim2018device} and in the current work do not provide a converse statement, i.e., 
		there might be schemes that do not use one-shot delivery and yield a load smaller than the bounds.  In addition, the converse bound in our paper considers the general 
		case of $K$ and $N$, whereas \cite{ibrahim2018device} only deals with the case where $K \leq N$. We also want to emphasize that, even under the constraint of one-shot delivery, handling the case 
		$K>N$ is more complicated than restricting to $K \leq N$ only \cite{ibrahim2018device}. 
		In fact, for $K>N$, the $K$ shared-link models are asymmetric in terms of users’ demands. 
		To derive our tight converse bound we leverage the connections among the $K$ shared-link caching models, 
		as we explain in Remark~\ref{rem:improvement}. \hfill $\lozenge$	 
		
	\end{remark}

	\section{A Novel Achievable D2D Coded Caching Scheme}\label{sec:achiev}
	
	In this section, we present a caching scheme that achieves the loads stated in Theorem \ref{teo} and Corollary \ref{corr}. To this end, we show that for any request vector $\boldsymbol{d}$ the proposed scheme achieves the load
	\begin{align}\label{eq:single}
		R^*(\boldsymbol{d},\boldsymbol{\boldsymbol{\mathcal{M}}_\textup{MAN}})=\frac{\binom{K-1}{t}-\frac{1}{K}\sum_{i=1}^{K}\binom{K-1-N_{\textup{e}}(\boldsymbol{d}_{\backslash\{i\}})}{t}}{\binom{K-1}{t-1}},
	\end{align}
	where $\boldsymbol{\mathcal{M}}_\textup{MAN}$ refers to the MAN symmetric placement. 
	This immediately proves the achievability of the average load in Theorem \ref{teo} and, together with the 
	explicit characterization of the worst-case demand in Appendix~\ref{sec:worstDemand}, also the achievability of 
	the worst-case load of Corollary \ref{corr}. 
	In the rest of this section we present the scheme and provide a simple example, illustrating how the idea 
	of exploiting common demands \cite{yu2018exact} is incorporated in the D2D setting. In Remark~\ref{rem:Ksharedlink}, 
	we will discuss our approach of decomposing the D2D model into $K$ shared-link models. 
	
	In the following, we restrict to integer values of $t \in [K]$. For non-integer values of $t$, the memory sharing idea 
	\cite{maddah2014fundamental,maddah2015decentralized,ji2016fundamental} can be routinely applied to achieve the lower convex envelope 
	of the achievable points for $t$ integer.
	
	\subsubsection{Placement phase}
	We use the MAN uncoded symmetric placement given in Definition \ref{def:MANplacement}.
	
	\subsubsection{Delivery phase}
	The delivery phase starts with the \emph{file-splitting} step: Each sub-file is divided into $t$ equal length disjoint sub-pieces of $\frac{F}{t\binom{K}{t}}$ bits which are denoted by $W_{q, \Vc, i}$, where $i\in \Vc$.
	
	Subsequently, each user $i$ selects any subset of $N_{\textup{e}}(\boldsymbol{d}_{\backslash\{i\}})$ users from $[K]\backslash\{i\}$, denoted by $\mathcal{U}^i=\{u_1^i,...,u^i_{N_{\textup{e}}(\boldsymbol{d}_{\backslash\{i\}})}\}$, which request $N_{\textup{e}}(\boldsymbol{d}_{\backslash\{i\}})$ distinct files. Extending the nomenclature in \cite{yu2018exact}, we refer to these users as \textit{leading demanders of user $i$}. 
	
	Let us now fix a user $i$ and consider an arbitrary subset $\mathcal{A}^i\subseteq [K]\backslash\{i\}$ of $t$ users. Each user $k\in \mathcal{A}^i$ needs the sub-piece $W_{d_k, \{\mathcal{A}^i \cup \{i\}\}\backslash\{k\}, i}$, which is cached by all the other users in $\mathcal{A}^i$ and at user $i$. 
	Precisely, all users $k \in \mathcal{A}^i$ shall retrieve the needed sub-pieces 
	$W_{d_k, \{\mathcal{A}^i \cup \{i\}\}\backslash\{k\}, i}$ from the transmissions of user $i$.
	By letting user $i$ broadcast the codeword 
	\begin{equation}\label{eq:Broadcasts}
	Y_{\mathcal{A}^i}^i:=\underset{k\in \mathcal{A}^i}{{\bigoplus}}W_{d_k, \{\mathcal{A}^i \cup \{i\}\}\backslash\{k\}, i},
	\end{equation} 
	this sub-piece retrieval can be accomplished, since each user $k \in \Ac^i$ has all the sub-pieces on the RHS 
	of \eqref{eq:Broadcasts}, except for $W_{d_k, \{\mathcal{A}^i \cup \{i\}\}\backslash\{k\}, i}$.
	
	We let each user $i$ broadcast (in sequence) the codewords $Y^i_{\mathcal{A}^i}$ for all subsets $\mathcal{A}^i$  such that 
	$\mathcal{A}^i\cap\mathcal{U}^i\neq\emptyset$, i.e., $X_i = \{Y^i_{\mathcal{A}^i}\}_{\mathcal{A}^i\cap\mathcal{U}^i\neq\emptyset}$. These are indeed all codewords
	that are directly useful for at least one leading demander of user $i$. 
	For each user $i \in [K]$, the size of the broadcasted codewords amounts to $\binom{K-1}{t}-\binom{K-1-N_{\textup{e}}(\boldsymbol{d}_{\backslash\{i\}})}{t}$ times 
	the size of a sub-piece, summing which for all $i \in [K]$ results in the load stated in \eqref{eq:single}.
	
	We now show that each user $k \in [K]$ is able to recover its needed sub-pieces.\footnote{Notice that some users may not be leading demanders of any other user, and the set of leading demander for different users may be different. Therefore, proving that {\em all} users can actually obtain {\em all} their needed bits is non-trivial.}
	When $k$ is a leading demander of a user $i$, i.e., $k\in \mathcal{U}^i$, it can decode any needed sub-piece $W_{d_k, \mathcal{B}^k \cup \{i\},i}$, where $\Bc^k \subseteq [K]\backslash\{i,k\}$, $|\Bc^k| = t-1$, from $Y^i_{\Bc^k \cup \{k\}}$ which is broadcasted from user $i$, by performing 
	
	\begin{equation}\label{eq:decodeBroadcast}
	W_{d_k, \mathcal{B}^k \cup \{i\},i} =  \left(\underset{x\in \mathcal{B}^k}{{\bigoplus}}W_{d_x, \{\mathcal{B}^k \cup \{i,k\}\}\backslash\{x\}, i}\right) \bigoplus Y^i_{\Bc^k\cup \{k\}}, 
	\end{equation}
	as it can be seen from \eqref{eq:Broadcasts}.
	
	However, when $k \notin \mathcal{U}^i$, not all of the corresponding codewords
	$Y^i_{\Bc^k \cup \{k\}}$ for its required sub-pieces $W_{d_k, \mathcal{B}^k\cup \{i\},i}$ are directly broadcasted from user $i$.
	In this case, user $k$ can still decode its needed sub-piece by generating the missing codewords based on its received codewords from user $i$. 
	To show this, we reformulate Lemma $1$ from \cite{yu2018exact}, applied to the codewords broadcasted by a user $i$.
	
	\begin{lemma}[Decodability]
		\label{dec_l}
		Given a user $i$, the request vector of the remaining users $\boldsymbol{d}_{\backslash\{i\}}$, and a set of leading demanders $\mathcal{U}^i$, for any subset $\mathcal{C}^i\subseteq [K]\backslash\{i\}$ that includes $\mathcal{U}^i$, let $\mathcal{V}^i_{\textup{F}}$ be family of  all subsets $\mathcal{V}^i$ of $\mathcal{C}^i$ such that each requested file in $\boldsymbol{d}_{\backslash\{i\}}$ is requested by exactly one user in $\mathcal{V}^i$.
		The following equation holds: 
		\begin{equation*}
			\underset{\mathcal{V}^i\in\mathcal{V}^i_{\textup{F}}}{\Large{\bigoplus}} Y^i_{\mathcal{C}^i\backslash\mathcal{V}^i}=0.
		\end{equation*}	
	\end{lemma}
	\begin{IEEEproof}
		As we will make it clear in Remark~\ref{rem:Ksharedlink}, the proposed scheme in fact corresponds to $K$ shared-link schemes. Thus, 
		Lemma~1 in \cite{yu2018exact} directly applies to each $i$-th  shared-link scheme.
	\end{IEEEproof}

	Let us now consider any subset $\mathcal{A}^i$ of $t$ non-leading demanders of user $i$
	such that 
	$\mathcal{A}^i\cap\mathcal{U}^i=\emptyset$.  
	Lemma \ref{dec_l} implies that the codeword $Y^i_{\mathcal{A}^i}$ can be directly computed from the broadcasted codewords by the following equation: 
	\begin{equation}\label{eq:lemmaRecNonleader}
	Y^i_{\mathcal{A}^i}=\underset{\mathcal{V}^i\in\mathcal{V}^i_{\textup{F}}\backslash \{\mathcal{U}^i\}}{\Large{\bigoplus}} Y_{\mathcal{C}^i\backslash\mathcal{V}^i},
	\end{equation}
	where $\mathcal{C}^i=\mathcal{A}^i\cup \,\mathcal{U}^i$, 
	because all codewords on the RHS of the above equation are directly broadcasted by user $i$. Thus, each user $k \notin \mathcal{U}^i$ can obtain the value $Y^i_{\mathcal{A}^i}$ for any subset $\mathcal{A}^i$ of $t$ users, and is able to decode its requested sub-piece.
	Hence, for each $i \in [K]\backslash\{k\}$, user $k$ decodes its needed sub-piece by following either one of the above strategies, 
	depending on whether it is a leading demander of $i$ or not.
	
	In the following, we provide an illustration of the above presented ideas through an example. 
	\paragraph*{Example}\label{para:ex}	
	Let us consider the case when $N=2, K=4, M=1, t=KM/N=2$ and $\boldsymbol{d} = (1,2,1,1)$. Notice that $N_{\textup{e}}(\boldsymbol{d}_{\backslash\{2\}})=1$ and $N_{\textup{e}}(\boldsymbol{d}_{\backslash\{i\}})=2\; \text{for}\,\, i \in \{1,3,4\}.$ Each file is divided into $\binom{4}{2} = 6$ sub-files and users cache the following sub-files for each $i \in \{1,2\}$:
	\begin{align*}
		\Mc_1 = \:\:\:& \lbrace W_{i,\{1,2\}}, W_{i,\{1,3\}},\, W_{i,\{1,4\}} \rbrace\\
		\Mc_2 = \:\:\:& \lbrace W_{i,\{1,2\}}, W_{i,\{2,3\}},\, W_{i,\{2,4\}} \rbrace\\
		\Mc_3 = \:\:\:& \lbrace W_{i,\{1,3\}}, W_{i,\{2,3\}},\, W_{i,\{3,4\}} \rbrace\\
		\Mc_4 = \:\:\:& \lbrace W_{i,\{1,4\}}, W_{i,\{2,4\}},\, W_{i,\{3,4\}} \rbrace 
	\end{align*}
	and need the following missing sub-files:
	\begin{align*}
		W_1 \backslash \Mc_1 = \:\:\:& \lbrace W_{1,\{2,3\}}, W_{1,\{2,4\}},\, W_{1,\{3,4\}} \rbrace \\
		W_2 \backslash \Mc_2 = \:\:\:& \lbrace W_{2,\{1,3\}}, W_{2,\{1,4\}},\, W_{2,\{3,4\}} \rbrace \\
		W_1 \backslash \Mc_3 = \:\:\:& \lbrace W_{1,\{1,2\}}, W_{1,\{1,4\}},\, W_{1,\{2,4\}} \rbrace \\
		W_1 \backslash \Mc_4 = \:\:\:& \lbrace W_{1,\{1,2\}}, W_{1,\{1,3\}},\, W_{1,\{2,3\}} \rbrace. 
	\end{align*}

	After splitting the sub-files into $2$ equal length sub-pieces, users $1,3,4$ transmit the following codewords, as can be seen from \eqref{eq:Broadcasts}:
\begin{align*}
X_1= \lbrace Y^1_{\{2,3\}} = \:\:\:& W_{2,\{1,3\},1} \Large{\oplus} W_{1,\{1,2\},1},\, Y^1_{\{2,4\}} = W_{2,\{1,4\},1} \Large{\oplus} W_{1,\{1,2\},1},\, Y^1_{\{3,4\}} = W_{1,\{1,3\},1} \Large{\oplus} W_{1,\{1,4\},1}\rbrace\\
X_3= \lbrace Y^3_{\{1,2\}} = \:\:\:& W_{1,\{2,3\},3} \Large{\oplus} W_{2,\{1,3\},3},\, Y^3_{\{1,4\}} = W_{1,\{1,3\},3} \Large{\oplus} W_{1,\{3,4\},3},\, Y^3_{\{2,4\}} = W_{2,\{3,4\},3} \Large{\oplus} W_{1,\{2,3\},3}\rbrace\\
X_4= \lbrace Y^4_{\{1,2\}} = \:\:\:& W_{1,\{2,4\},4} \Large{\oplus} W_{2,\{1,4\},4},\, Y^4_{\{1,3\}} = W_{1,\{1,4\},4} \Large{\oplus} W_{1,\{3,4\},4},\, Y^4_{\{2,3\}} = W_{2,\{3,4\},4} \Large{\oplus} W_{1,\{2,4\},4}\rbrace.
\end{align*}
	
	Notice that for these users, there exists no subset $\Ac^i$ s.t. $\Ac^i \subseteq [K]\backslash\{i\}$, $|\Ac^i|=t=2$ which satisfies $\Uc^i \cap \Ac^i \neq \emptyset$. However, depending on the choice of $\Uc^2$, user 2 can find $\binom{K-1-N_{\textup{e}}(\boldsymbol{d}_{\backslash\{2\}})}{t}=1$  subset $\Ac^2$ with $\Uc^2 \cap \Ac^2 \neq \emptyset$. Such an $\Ac^2$ can be determined as $\{3,4\}, \{1,4\}, \{1,3\}$ for the cases of $\Uc^2= \{1\}$, $\Uc^2 = \{3\}$, $\Uc^2 = \{4\}$, respectively.

	Picking user $1$ as its leading demander, i.e., $\mathcal{U}^2 = \{1\}$, user $2$ only transmits
\begin{align*}
X_2= \lbrace Y^2_{\{1,3\}} = \:\:\:& W_{1,\{1,2\},2} \Large{\oplus} W_{1,\{2,3\},2},\, Y^2_{\{1,4\}} = W_{1,\{1,2\},2} \Large{\oplus} W_{1,\{2,4\},2} \rbrace,
\end{align*}
	sparing the codeword $Y^2_{\{3,4\}} = W_{1,\{2,3\},2} \Large{\oplus} W_{1,\{2,4\},2}$. As mentioned before, the choice of the leading demanders is arbitrary and any one of the $Y^2_{\{1,3\}},\, Y^2_{\{1,4\}},\, Y^2_{\{3,4\}}$ can be determined as the redundant codeword. In fact, any one of these codewords can be attained by summing the other two, since $Y^2_{\{1,3\}} \Large{\oplus} Y^2_{\{1,4\}} \Large{\oplus} Y^2_{\{3,4\}} = 0$ (cf. \eqref{eq:lemmaRecNonleader}). 
	
	From the broadcasted codewords, all users can decode all their needed sub-pieces by using the sub-pieces in their caches as side-information, by performing \eqref{eq:decodeBroadcast}.
	
	As each sub-piece is composed of $F/t\binom{K}{t} = F/12$ bits and as $3 \times 3 + 1 \times 2 = 11$ codewords of such size are broadcasted, our scheme achieves a load of $11/12$, which could be directly calculated by \eqref{eq:single}.   \hfill $\lozenge$
	
	\begin{remark}\label{rem:Ksharedlink}
		Notice that a user $i$ generates its codewords exclusively from the sub-pieces $W_{q, \Vc, i}$ and there exist $\binom{K-1}{t-1}$ such sub-pieces in its cache.
		In addition, for any $k \in [K]\backslash\{i\}$, we have $W_{q, \Vc, i} \cap W_{q, \Bc, k} = \emptyset$ for any $\Vc, \Bc \subseteq [K]$, $|\Vc|=|\Bc|=t$, $i\in \Vc$, $k\in \Bc$. That is to say, users generate their codewords based on non-overlapping libraries of size $N\binom{K-1}{t-1}\frac{F}{t\binom{K}{t}} =  NF/K$ bits. 
		Also, observe that the cache of a user $k \neq i$ contains $\binom{K-2}{t-2}$ such $W_{q,\Vc,i}$ sub-pieces, which amounts to $N\binom{K-2}{t-2}\frac{F}{t\binom{K}{t}} =  \frac{N(t-1)F}{(K-1)K}$ bits. Recall that a sub-piece $W_{q,\Vc,i}$ is shared among $t-1$ users other than $i$.
		
		Therefore, the proposed scheme is in fact composed of $K$ shared-link models each with $N$ files of size $F' = F/K$ bits and $K' = K - 1$ users with caches of size $M' = \frac{N(t-1)}{(K-1)}$ units each. The corresponding parameter for each model is found to be $t' = \frac{K'M'}{N} = t-1$. Summing the loads of each $i \in [K]$ shared-link sub-systems \eqref{eq:singleSL} and replacing the shared-link system parameters  $F$, $K$, $M$, $t$, $N_{\textup{e}}(\boldsymbol{d})$ with
		$F'$, $K'$, $M'$, $t'$, and $N_{\textup{e}}(\boldsymbol{d}_{\backslash\{i\}})$, respectively, we obtain \eqref{eq:single}.  \hfill $\lozenge$
	\end{remark}
	
	\begin{remark}\label{rem:JiConnection}
		When each user requests a distinct file ($N_{\textup{e}}(\boldsymbol{d})=K$), our proposed scheme corresponds to the one originally presented in \cite{ji2016fundamental}. The potential improvement of our scheme when $N_{\textup{e}}(\boldsymbol{d}) < K$ hinges on identifying the possible linear dependencies among the codewords generated by a user. 
		\hfill $\lozenge$
	\end{remark}
	
	\begin{remark}
		\label{rem:symmetry in file-splitting}
		The presented scheme is symmetric in the placement phase and in the file-splitting step in the delivery phase. 
		The optimality of the symmetry in placement phase \cite{maddah2014fundamental} was already shown for the shared-link model 
		in \cite{wan2016caching,wan2016optimality,yu2018exact} under the constraint of uncoded placement (and random uniform demands, for the average load case). 
		This symmetry is intuitively plausible as the placement phase takes place before users reveal their demands and any asymmetry in the placement definitely 
		would not lead to a better worst-case load.
		However, a file-splitting step occurs after users make their demands known to the other users. Interestingly, it turns out that the proposed caching scheme with symmetric file-splitting step achieves the lower bound shown in Section \ref{sec:Converse}, 
		even though the $K$ shared-link sub-systems may not have the same value 
		of $N_{\textup{e}}(\boldsymbol{d}_{\backslash\{i\}})$ and are therefore {\em asymmetric}, given the request vector.
	\end{remark}

	\section{Load-outage trade-off in the presence of random user activity}\label{sec:loadouta}
	
	In this section we consider the more realistic scenario of random user activity (again focusing on $t = \frac{KM}{N} \in [K]$). 
	We assume that each user might be inactive independently with a probability of $p$ during the delivery phase and that  $I$ 
	is the realization of the number of inactive users. We also assume that  the inactivity event of one user is not known by the other users 
	during the delivery phase.
	
	To tackle the presence of such inactive users, we use MDS coding. 
	A codeword encoded using an $(m,n)$-MDS code can be perfectly reconstructed from any $m$ out of $n$ 
	MDS-coded blocks, with the penalty of an increase in the block length by a factor of $n/m$. 
	We define the system outage probability $P_{out}$ as the probability that there exists some active user who can not decode their desired file.
	Notice that this is different from the per-user outage probability, considered for example in \cite{ji2016fundamental}. 
	
	Towards this end, we recall that for the D2D caching problem in~\cite{ji2016fundamental},   our proposed caching scheme in Section~\ref{sec:achiev} divides each file 
	into $t \binom{K}{t}$ sub-pieces and let each user cache $t \binom{K-1}{t-1}$ sub-pieces of each file. 
	The proposed one-shot delivery scheme allows users to decode exactly $\binom{K-2}{t-1}$ needed sub-pieces of their requested file from each of the other $K-1$ users. 
	However, if user inactivity exists, each active user cannot receive all the $(K-1)\binom{K-2}{t-1}$ needed sub-pieces. 
	Instead, each active user can only receive totally $(K-I-1)\binom{K-2}{t-1}$ sub-pieces from the remaining $K-I-1$ active 
	users.
	
	Notice that the user inactivity event happens during the delivery phase which can not be predicted in prior during the placement phase.   
	In order to cope with random inactivity, we fix an integer $a\in [0:K-1]$ and divide each file into $t \binom{K-1}{t-1} + (K-1-a)\binom{K-2}{t-1}$ 
	non-overlapping and equal-length parts, which are then encoded by a $(t \binom{K-1}{t-1} + (K-1-a)\binom{K-2}{t-1},t \binom{K}{t})$-MDS code. 
	Hence,  for each file $W_i$ where $i\in [N]$, we have $t \binom{K}{t}$ coded sub-pieces, each of which has length $\frac{F}{t \binom{K-1}{t-1} + (K-1-a)\binom{K-2}{t-1}}$ bits. 
	For each set $\Vc\subseteq [K]$ where $|\Vc|=t+1$ and each user $k\in \Vc$, there is one coded sub-piece $W^{\prime}_{i,\Vc,k}$ 
	cached by users in $\Vc$.
	
	During the delivery phase, we use the one-shot delivery scheme in Section~\ref{sec:achiev}.	
	Hence, an active user caches $t \binom{K-1}{t-1} $ coded sub-pieces of its desired file and receives $(K-1-I)\binom{K-2}{t-1}$ coded sub-pieces  of its desired 
	file during the delivery phase from other active users. It can be seen that if $a\geq I$, each active user can recover its desired file using MDS decoding. 
	However, larger values of $a$ yield an increase of the load and of the required cache size, both by a factor  of $\frac{t \binom{K}{t}}{t \binom{K-1}{t-1} + (K-1-a)\binom{K-2}{t-1}}$. 
	Hence, there exists a tradeoff between the load and cache memory and the system outage probability. 	
	For each $a\in [0:K-1]$, the outage probability of the network is given by the binomial tail	probability
	\begin{align}
		P_{out} = \mathbb{P} \{a < I\}= \sum_{i^{\prime} = a+1}^{K} \binom{K}{i^{\prime} }p^{i^{\prime} } (1-p)^{K-i^{\prime} }.\label{eq:outage}
	\end{align}
	Hence, we have the following results.
	\begin{theorem}[Average load with user inactivity]\label{thm:inact}
		For a D2D caching problem  as defined in Theorem \ref{teo} with user inactivity probability $p$, 
		the following memory and average load tradeoff points with uniform distribution are achievable
		\begin{equation}
		 (M_{\textup{inact}},R_{\textup{ave,inact}})=\Big( \frac{t \binom{K}{t}}{t \binom{K-1}{t-1} + (K-1-a)\binom{K-2}{t-1}}M, \frac{t \binom{K}{t}}{t \binom{K-1}{t-1} + (K-1-a)\binom{K-2}{t-1}}R^*_{\textup{ave, o}}\Big), \label{eq:averageinact}
		\end{equation}
		with an  outage probability in~\eqref{eq:outage} for each $a\in[0:K-1]$ and each $t\in [0:K]$, 
		where 	$R^*_{\textup{ave, o}}$ is given in~\eqref{eq:averageworstcase}.
		For other memory size, the memory and average load tradeoff  point can be obtained by the lower convex envelope for the above corner points.  \hfill $\square$
	\end{theorem}

	\begin{theorem}[Worst-case load with user inactivity]\label{thm:worst inact}
		For a D2D caching problem  as defined in Theorem \ref{teo} with user inactivity probability $p$, 
		the following memory and worst-case load tradeoff points, are achievable
		\begin{equation}
		 (M_{\textup{inact}},R_{\textup{worst,inact}})	= \Big( \frac{t \binom{K}{t}}{t \binom{K-1}{t-1} + (K-1-a)\binom{K-2}{t-1}}M, \frac{t \binom{K}{t}}{t \binom{K-1}{t-1} + (K-1-a)\binom{K-2}{t-1}}R^*_{\textup{worst, o}}\Big), \label{eq:worstcaseinact}
		\end{equation}
		with an  outage probability in~\eqref{eq:outage} for each $a\in[0:K-1]$ and each $t\in [0:K]$, 
		where $R^*_{\textup{worst, o}}$ is given in~\eqref{eq:worstCaseload}.
		For other memory size, the  memory and worst-case load tradeoff point can be obtained by the lower convex envelope for the above corner points.
		\hfill $\square$
	\end{theorem}

	Notice that if we do not use a one-shot delivery scheme,  each user will decode some of its needed sub-pieces by the received packets from several other users. 
	Since the inactivity events of other users are not known by each user, such non one-shot delivery scheme would be hard to design and definitely not intuitive. 
	
	\begin{remark}
		The user inactivity problem in  D2D caching setting was previously discussed in \cite{TebbiSung}. However, the solution proposed there 
		requires the knowledge of the inactive users, which is a highly impractical assumption for a D2D network where users may turn on and off
		in an independent and decentralized manner.  Moreover, the proposed scheme can also deal with the straggler problem, where the  
		the ending time of the transmission by each user is different and unexpected in prior. From the proposed delivery scheme, 
		  After  the  successful  transmissions of any  $K-a$ users,  each user is able to recover its desired file, and thus
		  we could spare the waiting times of   $a$ lowest users' transmissions. 
		
		\hfill $\lozenge$
	\end{remark}	
	
	\section{Numerical Evaluations}\label{sec:Numerical}

	In this section, we compare the load-memory trade-off of the presented scheme with the bounds from related works and evaluate the load-outage trade-off when user inactivity is taken into account (cf. Section \ref{sec:loadouta}).
	
\begin{figure}[ht]
		{\begin{minipage}[t]{250pt}
				\hspace{-5pt}	\includegraphics[scale=0.52,trim=1cm 0cm 1.5cm 0.5cm,clip]{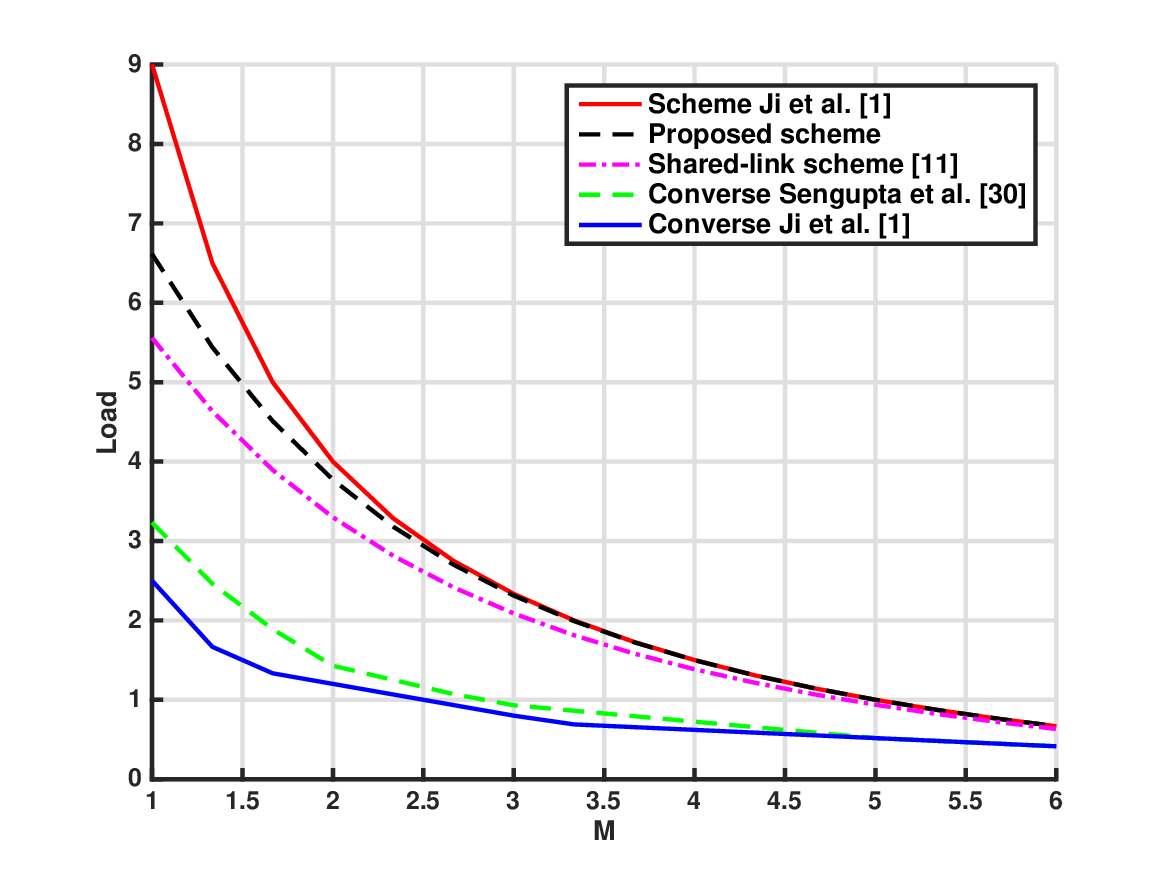}
			\end{minipage}}
			\hfill
			{\begin{minipage}[t]{250pt}
					\hspace{-5pt}	{\includegraphics[scale=0.52,trim=1cm 0cm 1.5cm 0.5cm,clip]{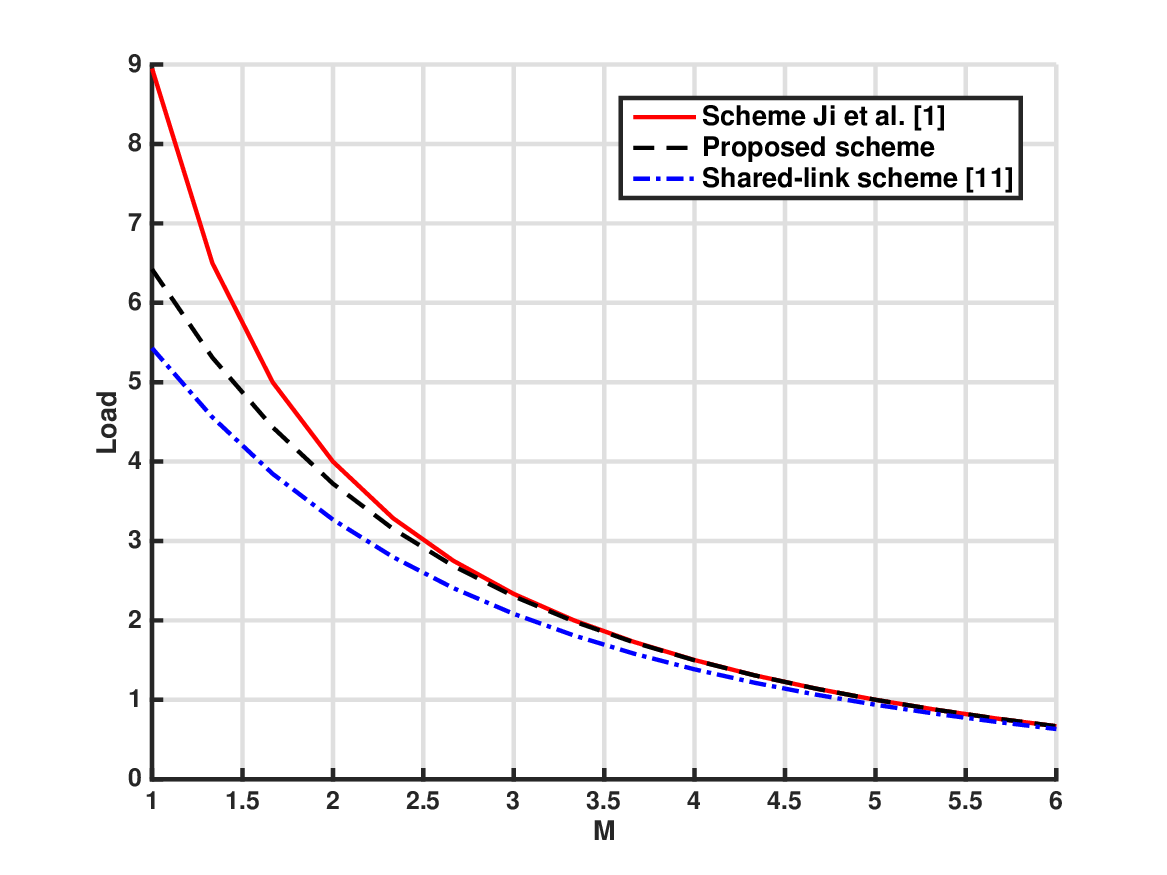}}
				\end{minipage}}
				\caption{\small Consider the D2D caching problem in~\cite{ji2016fundamental} with $N=10$, and $K=30$.  The figure on the left is for  the tradeoff between memory size and the worst-case load. The figure on the right shows the tradeoff between memory size and the average load with uniform demand distribution.  }
				\label{fig:N10K20}
			\end{figure}

	In Fig.~\ref{fig:N10K20}, we consider the D2D caching problem in~\cite{ji2016fundamental} with $N=10$, $K=30$ and compare the load achieved by the presented one-shot scheme with the achievable load in \cite{ji2016fundamental} (cf. Subsection \ref{sub:D2DRelated}) and with the minimum achievable load for the shared-link model \cite{yu2018exact} (cf. Subsection \ref{sub:sharedLinkRelated}). When the minimum worst-case load is considered, we also provide the lower bounds in \cite{ji2016fundamental,sengupta2015beyond}. It can be seen that the proposed D2D caching scheme outperforms the one in~\cite{ji2016fundamental}.

			\begin{figure}[ht]
				{\begin{minipage}[t]{250pt}
						\hspace{-5pt}	\includegraphics[scale=0.50,trim=1cm 0cm 1.5cm 0.5cm,clip]{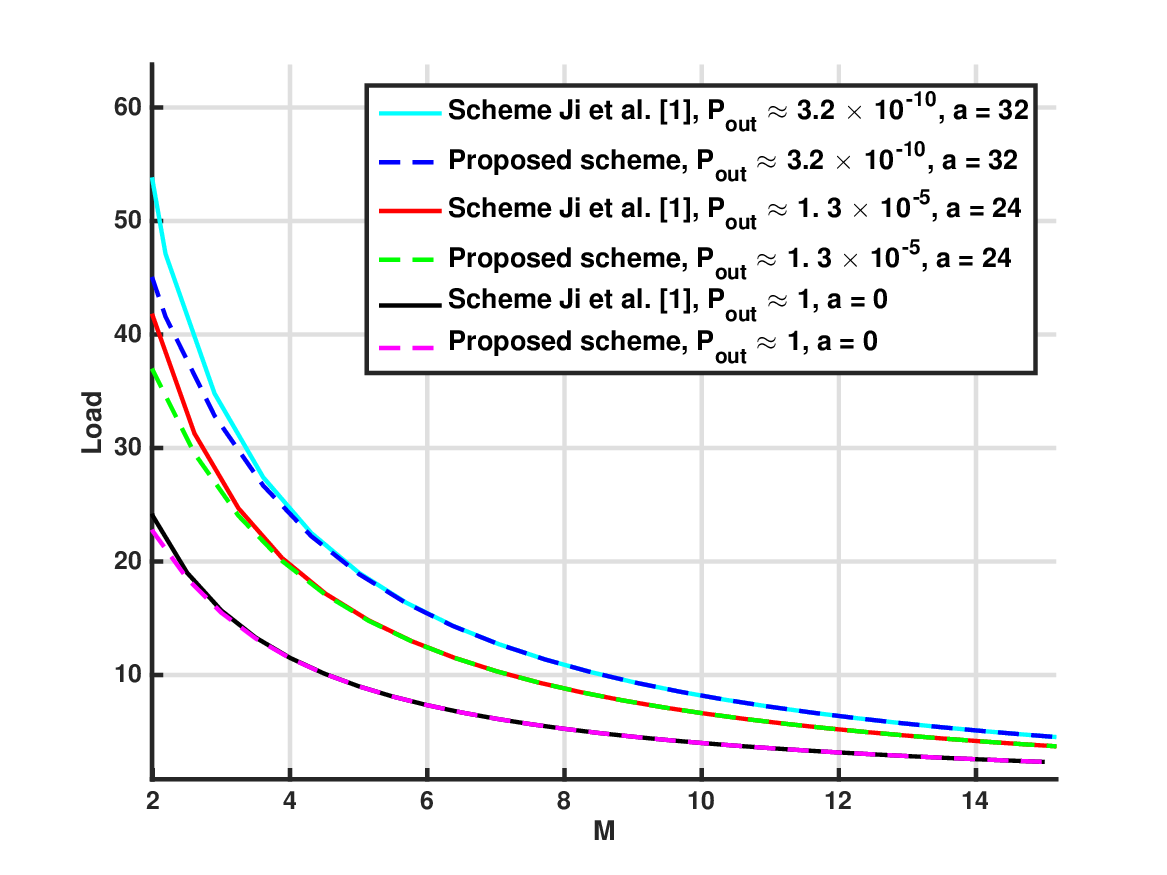}
					\end{minipage}}
					\hfill
					{\begin{minipage}[t]{250pt}
							\hspace{-5pt}		\includegraphics[scale=0.50,trim=1cm 0cm 1.5cm 0.5cm,clip]{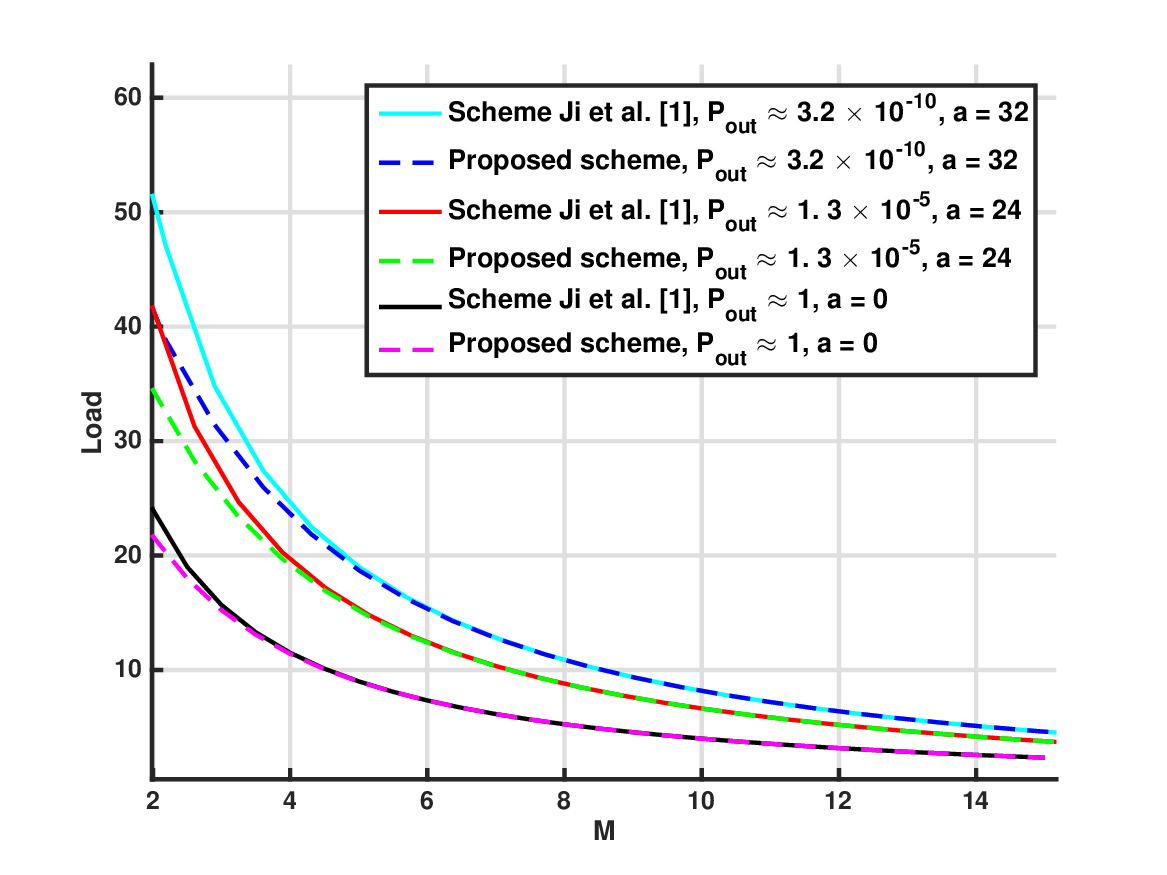}
						\end{minipage}}
						\caption{\small Consider the D2D caching problem  with user inactivity, where    $N=50$, $K=100$, and $p=0.1$.  The figure on the left is for  the tradeoff between memory size and the worst-case load. The figure on the right shows the tradeoff between memory size and the average load with uniform demand distribution.  }
						\label{fig:loadCacheOutage51015}
					\end{figure}

					In Fig.~\ref{fig:loadCacheOutage51015}, we consider the D2D caching problem with user inactivity, where    $N=50$, $K=100$, and $p=0.1$.  We see that for small cache sizes increasing the load by a factor of $2$ is sufficient to drive the outage probability to $P_{out}\approx 10^{-5}$. As the cache size grows, the increase in the load necessary to achieve this outage diminishes.
					

					
					\section{Conclusions}	
					In this work, we completely characterized the  memory-load
					trade-off for cache-aided D2D networks under the constraint of one-shot delivery, when the placement phase is restricted to be uncoded and centralized. We presented a caching scheme and
					proved its exact optimality in terms of both average load under uniformly distributed user demands 
					and worst-case load. Furthermore, we showed that the achieved load is optimal within a factor of $2$  under the constraint of uncoded cache placement, and within a factor of $4$ in general.  Lastly, we extended the proposed scheme to the case of MDS coding 
					in order to guarantee a desired level of  robustness against random user inactivity.
					
					Although the information theoretic coded caching model is admittedly very idealized, progress has been made to apply such ideas in practical scenarios. 
					It is clear that the ability of performing the placement phase off-line, on a much slower time scale than the delivery phase, is crucial to this setting. 
					For example, one may consider the cased of ``Bring Your Own'' (BYO) device inflight entertainment systems, where some hundreds of users are simultaneously located 
					in a small space (a bus, a train car, or an aircraft). Suppose that users preloaded an App, that preloads pieces of a content library from their home Internet to their devices. 
					Then, a scenario where each device can cache (say) $\sim 10$ movies, and a total library of (say) $\sim 100$ movies, with (say) $\sim 200$ users, 
					is not far from reality. For such a scenario, the users could collectively access to the whole library in an on-demand fashion, using schemes inspired by the D2D coded caching scheme
					presented in this paper. 
					
					\appendices
					
					\section{Converse Bound under the Constraint of  One-Shot Delivery}\label{sec:Converse}
					In this section we propose the converse bound under the constraint of one-shot delivery given in Theorem~\ref{teo}. Under the constraint of one-shot delivery, we can divide each sub-file $W_{i,\Vc}$ into sub-pieces. Recall that $W^{k,i}_{d_k,\Vc}$ represents the bits of $W_{d_k}$ decoded by user $k$ from $X_{i}$. 
					Under the constraint of one-shot delivery, we can divide the D2D caching problem into $K$ shared-link models. In  the $i^{\textrm{th}}$ shared-link model where $i\in [K]$, user $i$  transmits $X_i$ such that each user $k\in [K]\setminus \{i\}$ can recover $W^{k,i}_{d_k,\Vc}$  for all $\Vc\subseteq ([K]\setminus \{k\})$ where $i\in \Vc$.

					\subsection{Converse Bound for $R^*_{\textup{o}}(\mathbf{d},\boldsymbol{\mathcal{M}})$}
					\label{sub:converse for R(d,M)}
					Fix a request vector $\mathbf{d}$ and a cache placement $\boldsymbol{\mathcal{M}}$.  We first focus on the shared-link model where user $i\in[K]$ broadcasts.
					
					Consider a permutation of $[K]\setminus \{i\}$, denoted by $\mathbf{u}=(u_{1},u_{2},...,u_{K-1})$.
					For given permuted users $\mathbf{u}$ and request $\mathbf{d}$ vectors, 
					we define a new vector $\mathbf{f}(\mathbf{u},\mathbf{d})$ obtained by successive pruning of the vector $\mathbf{u}$
					by iterating the following steps: let $\mathbf{f}^0= \mathbf{u}$ (initial state), and for each  $\ell =1,2,\ldots$, 
					let $\mathbf{f}^{\ell}$ be the vector obtained from $\mathbf{f}^{\ell-1}$ by removing all elements $f^{\ell-1}_j$ (the $j^{\textrm{th}}$ element of $\mathbf{f}^{\ell-1}$)  with $ j > \ell$ such that $d_{f^{\ell-1}_j} = d_{f^{\ell-1}_{\ell}}$. We
					stop when there are no more elements to remove, and call the resulting vector $\mathbf{f}(\mathbf{u},\mathbf{d})$. 
					In other words, $\mathbf{f}(\mathbf{u},\mathbf{d})$ is obtained from $\mathbf{u}$ and $\mathbf{d}$ by removing from $\mathbf{u}$, for each demanded file, all the users demanding 
					such file except the user in the leftmost position of $\mathbf{u}$. 
					For example, if $\mathbf{u}=(2,3,5,4)$, $\mathbf{d}=(1,2,2,3,3)$, we have $d_{u_1}=d_{u_2}=2$ and $d_{u_3}=d_{u_4}=3$, and thus $\mathbf{f}(\mathbf{u},\mathbf{d})=(u_1, u_3)=(2,5)$. It can be seen that $\mathbf{f}(\mathbf{u},\mathbf{d})$ contains $N_{\textup{e}}(\boldsymbol{d}_{\backslash\{i\}})$ elements.
					For each $j\in [1:N_{\textup{e}}(\boldsymbol{d}_{\backslash\{i\}})]$, we define $f_j(\mathbf{u},\mathbf{d})$ as the $j^{\textrm{th}}$ element of $\mathbf{f}(\mathbf{u},\mathbf{d})$.
					
					For the permutation $\mathbf{u}$, we can choose a set of sub-pieces, $\big(W^{f_j(\mathbf{u},\mathbf{d}),i}_{d_{f_j(\mathbf{u},\mathbf{d})},\Vc_{j}} : 
					\Vc_{j}\subseteq[K]\backslash \{f_1(\mathbf{u},\mathbf{d}),\ldots,f_j(\mathbf{u},\mathbf{d})\}, \ i\in \Vc_j,
					\ j\in[N_{\textup{e}}(\boldsymbol{d}_{\backslash\{i\}})]
					\big)$. 
					By a similar proof as~\cite[Lemma 1]{wan2016optimality} (as used in Section~\ref{sub:shared-link converse} of this paper), we have the following lemma.
					\begin{lemma}
						\label{lem:acyclic}
						For each permutation of $[K]\setminus \{i\}$, denoted by $\mathbf{u}=(u_{1},u_{2},...,u_{K-1})$, we have
						\begin{align}
							H(X_i)
							&\geq \sum_{j\in[N_{\textup{e}}(\boldsymbol{d}_{\backslash\{i\}})]} \sum_{ \Vc_{j}\subseteq[K]\backslash \{f_1(\mathbf{u},\mathbf{d}),\ldots,f_j(\mathbf{u},\mathbf{d})\}: i\in \Vc_j}   |W^{f_j(\mathbf{u},\mathbf{d}),i}_{d_{f_i(\mathbf{u},\mathbf{d})},\Vc_{j}}|.
							\label{eq:converse of Xk}
						\end{align}
					\end{lemma} 
					\begin{IEEEproof}
						In the $i^{\textrm{th}}$ shared-link model, for each $\mathbf{u}$ which is a permutation of $[K]\setminus \{i\}$, we can generate a directed graph. Each sub-piece $W^{f_j(\mathbf{u},\mathbf{d}),i}_{d_{f_j(\mathbf{u},\mathbf{d})},\Vc_{j}}$ where $j\in[N_{\textup{e}}(\boldsymbol{d}_{\backslash\{i\}})]$, $\Vc_{j}\subseteq[K]\backslash \{f_1(\mathbf{u},\mathbf{d}),\ldots,f_j(\mathbf{u},\mathbf{d})\}$ and  $i\in \Vc_j$, is represented by an independent node in the graph demanded by user $f_j(\mathbf{u},\mathbf{d})$. There is a directed edge from node $j_1$ to node $j_2$, if and
						only if the user who demands the sub-piece  represented by node $j_2$ caches the sub-piece represented
						by node $j_1$.

						We say that sub-pieces $W^{f_j(\mathbf{u},\mathbf{d}),i}_{d_{f_j(\mathbf{u},\mathbf{d})},\Vc_{j}}$ for all  $\Vc_{j}\subseteq[K]\backslash \{f_1(\mathbf{u},\mathbf{d}),\ldots,f_j(\mathbf{u},\mathbf{d})\}$ where  $i\in \Vc_j$, are in level $j$.
						It is easy to see that the user demanding each sub-piece in level $j$  
						knows
						neither the sub-pieces in the same level, nor the sub-pieces in the
						higher levels. As a result, 
						in the directed graph, there is no sub-set containing a directed cycle. Hence, by the acyclic index coding converse bound in~\cite[Corollary 1]{onthecapacityindex}, we have~\eqref{eq:converse of Xk}.
					\end{IEEEproof}

					Considering all the permutations of $[K]\setminus \{i\}$ and all $i\in [K]$, we sum the inequalities in form of~\eqref{eq:converse of Xk} to obtain,
					\begin{equation}
						(K-1)! \big(H(X_1)+\ldots+H(X_K)  \big) \geq\sum_{k\in[K]} \sum_{\Vc\subseteq [K]\setminus \{k\}} \sum_{i\in \Vc} a^{k,i}_{\Vc} |W^{k,i}_{d_k,\Vc}|, \label{eq:summing all Xk}
					\end{equation}
					where $a^{k,i}_{\Vc}$  represents the coefficient of $|W^{k,i}_{d_k,\Vc}|$ in the sum. In Appendix~\ref{sec:proof of same coeff}, we prove the following lemma.
					\begin{lemma}
						\label{lem:same coeff}
						$a^{k,i_1}_{\Vc}=a^{k,i_2}_{\Vc}$, for each $i_1,i_2 \in \Vc$.
					\end{lemma}
					
					From Lemma~\ref{lem:same coeff}, we define $a^{k}_{\Vc}=\frac{ a^{k,i}_{\Vc}}{(K-1)! } $ for all $i\in \Vc$. Hence, from~\eqref{eq:summing all Xk} we have
\begin{subequations} 
	\begin{align}
	R^*_{\textup{o}}(\mathbf{d},\boldsymbol{\mathcal{M}})F\geq 
	\big(H(X_1)+\ldots+H(X_K) \big) &\geq \frac{1}{(K-1)! } \sum_{k\in[K]} \sum_{\Vc\subseteq [K]\setminus \{k\}} \sum_{i\in \Vc} a^{k,i}_{\Vc} |W^{k,i}_{d_k,\Vc}| \label{eq:sum to sub-file 1} \\
	& \geq \sum_{k\in[K]} \sum_{\Vc\subseteq [K]\setminus \{k\}} a^{k}_{\Vc} |W_{d_k,\Vc}|\label{eq:sum to sub-file}
	\end{align} 
\end{subequations} 
					where in~\eqref{eq:sum to sub-file} we used
					\begin{align}
						\sum_{i\in \Vc}|W^{k,i}_{d_k,\Vc}|\geq |W_{d_k,\Vc}|.\label{eq:sum to sub-file 3}\
					\end{align}
					
					\begin{remark}\label{rem:improvement}
						To derive the converse bound under the constraint of uncoded cache placement in~\cite{yu2018exact,wan2016caching},  the authors consider all the demands and all the permutations and sum the inequalities together. By the symmetry, it can be easily checked that in the summation expression, the coefficient of sub-files known by the same number of users is the same. However, in our problem, notice that~\eqref{eq:sum to sub-file 1} and~\eqref{eq:sum to sub-file 3} only hold for one demand. So each time we should consider one demand and let the coefficients of $H(X_k)$ where $k\in[K]$ be the same. Meanwhile, for each demand, we also should let the coefficients in Lemma~\ref{lem:same coeff} be the same.  
						However, for each demand, the $K$ shared-link models are not symmetric. If we use the choice of the acyclic sets   in~\cite{yu2018exact,wan2016caching} for each of the $K$ shared-link models,  we cannot ensure that for one demand, the coefficients are    symmetric. 
					\end{remark}
					\subsection{Converse Bound for $R^*_{\textup{ave, o}}$}
					\label{sub:average converse}
					As in \cite{tian2016symmetry,yu2018exact}, we group the request vectors in $\mathcal{D}$ according to the frequency of common entries that they have. Towards this end, for a request $\boldsymbol{d}$, we stack in a vector of length $N$ the number of appearances of each request in descending order, and denote it by $\boldsymbol{s}(\boldsymbol{d})$. 
					We refer to this vector as \textit{composition} of $\boldsymbol{d}$. Clearly, $\sum_{n=1}^{N}s_n(\boldsymbol{d})=K$.  By $\mathcal{S}$ we denote the set of all possible compositions. We denote the set of request vectors with the same composition $\boldsymbol{s} \in \mathcal{S}$  by $\mathcal{D}_{\boldsymbol{s}}$. We refer to these subsets as \textit{type}s. Obviously, they are disjoint and $\bigcup\limits_{\boldsymbol{s}\in \mathcal{S}} \mathcal{D}_{\boldsymbol{s}} = \mathcal{D}$. For instance, for $N=3$ and $K=5$, one has $\mathcal{S} = \{(5,0,0), (4,1,0), (3,2,0), (3,1,1), (2,2,1)\}$ and $\mathcal{D}_{\boldsymbol{s}} = \{(1,1,1,1,1), (2,2,2,2,2), (3,3,3,3,3)\}$ when $\boldsymbol{s} = (5,0,0)$.
					
					We focus on a type of requests $\boldsymbol{s}$. For each request vector $\mathbf{d} \in \Dc_{\boldsymbol{s}}$, we lower bound $R^*_{\textup{o}}(\mathbf{d},\boldsymbol{\mathcal{M}})$ as~\eqref{eq:sum to sub-file}. Considering all the requests in $\Dc_{\boldsymbol{s}}$, we then sum the inequalities in form of~\eqref{eq:sum to sub-file},
					\begin{align}
						\sum_{\mathbf{d} \in \Dc_{\boldsymbol{s}}} R^*_{\textup{o}}(\mathbf{d},\boldsymbol{\mathcal{M}})F\geq \sum_{q\in[N]} \sum_{\Vc\subseteq [K]} b_{q,\Vc} |W_{q,\Vc}|\label{eq:demand type}
					\end{align}
					where $b_{q,\Vc}$ represents the coefficient of $|W_{q,\Vc}|$. By the symmetry, it can be seen that $b_{q_1,\Vc_1}=b_{q_2,\Vc_2}$ if $|\Vc_1|=|\Vc_2|$. So we let $b_{t}:=b_{q,\Vc} $ for each $q\in [N]$ and $\Vc\subseteq [K]$ where $|\Vc|=t$. Hence, from~\eqref{eq:demand type} we get 
					\begin{align}
						|\Dc_{\boldsymbol{s}}|F \mathbb{E}_{\boldsymbol{d}\in \Dc_{\boldsymbol{s}}}[ R^*_{\textup{o}}(\boldsymbol{d},\boldsymbol{\mathcal{M}})]=\sum_{\mathbf{d} \in \Dc_{\boldsymbol{s}}} R^*_{\textup{o}}(\mathbf{d},\boldsymbol{\mathcal{M}})F\geq \sum_{t\in [0:K]} b_{t} x_t \label{eq:bt xt}
					\end{align}
					where we define
					\begin{align}
						x_t:= \sum_{q\in [N]} \sum_{\Vc \subseteq [K]:|\Vc|=t} |W_{q,\Vc}|.\label{eq:definition of xt}
					\end{align}
					Notice that each sub-file (assumed to be $W_{q,\Vc}$) demanded by each user is transmitted as $|\Vc|$ sub-pieces, each of which is known by $|\Vc|$ users. Now focus on an integer $t\in [0;K]$. We compute $b_{t}$ in the next two steps:
					\begin{enumerate}
						\item  
						$x_t$ is the sum of $N\binom{K}{t}$ sub-files known by $t$ users.
						So in the sum expression~\eqref{eq:bt xt}, we obtain $b_{t} x_t$ from a sum of $tN\binom{K}{t}b_t$ terms of sub-pieces known by $t$ users.
						
						\item 
						In~\eqref{eq:converse of Xk}, there are $\binom{K-2}{t-1}+\binom{K-3}{t-1}+\dots+\binom{K-N_{\textup{e}}(\boldsymbol{d}_{\backslash\{k\}})-1}{t-1}=\binom{K-1}{t}-\binom{K-N_{\textup{e}}(\boldsymbol{d}_{\backslash\{k\}})-1}{t}$, terms of sub-pieces known by $t$ users. Hence,~\eqref{eq:sum to sub-file 1} contains $\sum_{i\in [K]} \binom{K-1}{t}-\binom{K-N_{\textup{e}}(\boldsymbol{d}_{\backslash\{i\}})-1}{t}$ terms of sub-pieces known by $t$ users. So the sum of~\eqref{eq:sum to sub-file 1}  over all the $|\Dc_{\boldsymbol{s}}|$ request vectors, contains $|\Dc_{\boldsymbol{s}}| \Big( \sum_{i\in [K]} \binom{K-1}{t}-\binom{K-N_{\textup{e}}(\boldsymbol{d}_{\backslash\{i\}})-1}{t}\Big)$ terms of sub-pieces known by $t$ users.  
					\end{enumerate}
					Combining Steps 1) and 2), we can see 
					\begin{align}
						tN\binom{K}{t}b_t= |\Dc_{\boldsymbol{s}}| \left( \sum_{i\in [K]} \binom{K-1}{t}-\binom{K-N_{\textup{e}}(\boldsymbol{d}_{\backslash\{i\}})-1}{t}\right)\label{eq:step 1 2}
					\end{align}
					and thus 
					\begin{align}
						b_t=\frac{|\Dc_{\boldsymbol{s}}| \Big( \sum_{i\in [K]} \binom{K-1}{t}-\binom{K-N_{\textup{e}}(\boldsymbol{d}_{\backslash\{i\}})-1}{t}\Big)}{tN\binom{K}{t}}.\label{eq:bt}
					\end{align}
					We take~\eqref{eq:bt} into~\eqref{eq:bt xt} to obtain
					\begin{subequations} 
						\begin{align}
							&\mathbb{E}_{\boldsymbol{d}\in \Dc_{\boldsymbol{s}}}[ R^*_{\textup{o}}(\boldsymbol{d},\boldsymbol{\mathcal{M}})] \geq \sum_{t\in [0:K]} \frac{b_{t} x_t}{|\Dc_{\boldsymbol{s}}|F} \label{eq:take bt into xt 1}\\
							&= \sum_{t\in [0:K]} \frac{\Big( \sum_{i\in [K]} \binom{K-1}{t}-\binom{K-N_{\textup{e}}(\boldsymbol{d}_{\backslash\{i\}})-1}{t}\Big) x_t}{tN\binom{K}{t} F} \label{eq:take bt into xt 2}\\
							&=\sum_{t\in [0:K]} \frac{\Big( \sum_{i\in [K]} \binom{K-1}{t}-\binom{K-N_{\textup{e}}(\boldsymbol{d}_{\backslash\{i\}})-1}{t}\Big) x_t}{K\binom{K-1}{t-1} NF} \label{eq:take bt into xt 3}\\
							&=\sum_{t\in [0:K]} \frac{ \binom{K-1}{t}- \frac{1}{K} \sum_{i\in [K]}\binom{K-N_{\textup{e}}(\boldsymbol{d}_{\backslash\{i\}})-1}{t} }{\binom{K-1}{t-1} NF}x_t \label{eq:take bt into xt 4}.
						\end{align}
					\end{subequations} 
					We also have the constraint of file size and 
					\begin{align}
						\sum_{t\in [0:K]} x_t=NF, \label{eq:file size}
					\end{align}
					and the constraint of cache size 
					\begin{align}
						\sum_{t\in [1:K]} t x_t \leq KMF. \label{eq:cache size}
					\end{align}
					Note that the set of values of $N_{\textup{e}}(\boldsymbol{d}_{\backslash\{i\}})$ for all $i \in [K]$ is the same for all request vectors $\boldsymbol{d}$ with given composition $\boldsymbol{s}$. We let $r_{t,\boldsymbol{s}}:=\frac{ \binom{K-1}{t}- \frac{1}{K} \sum_{i\in [K]}\binom{K-N_{\textup{e}}(\boldsymbol{d}_{\backslash\{i\}})-1}{t} }{\binom{K-1}{t-1} NF}$, where $\boldsymbol{d}\in \Dc_{\boldsymbol{s}}$. Similar to~\cite{yu2018exact}, we can lower bound~\eqref{eq:take bt into xt 4} using Jensen's inequality and the
					monotonicity of $\textrm{Conv}(r_{t,\boldsymbol{s}})$,
					\begin{align}
						\mathbb{E}_{\boldsymbol{d}\in \Dc_{\boldsymbol{s}}}[ R^*_{\textup{o}}(\boldsymbol{d},\boldsymbol{\mathcal{M}})]\geq \textrm{Conv}(r_{t,\boldsymbol{s}}).
					\end{align}
					So we have
					\begin{align}
						\min_{\boldsymbol{\mathcal{M}}}\mathbb{E}_{\boldsymbol{d}\in \Dc_{\boldsymbol{s}}}[ R^*_{\textup{o}}(\boldsymbol{d},\boldsymbol{\mathcal{M}})]&\geq \min_{\boldsymbol{\mathcal{M}}} \textrm{Conv}(r_{t,\boldsymbol{s}})=\textrm{Conv}(r_{t,\boldsymbol{s}}).\label{eq:conv}
					\end{align}
					Considering all the request types and from~\eqref{eq:conv}, we have
					\begin{align}
						R^*_{\textup{ave, o}}&\geq \mathbb{E}_{\boldsymbol{s}}\left[ \min_{\boldsymbol{\mathcal{M}}} \mathbb{E}_{\boldsymbol{d}\in \Dc_{\boldsymbol{s}}}[ R^*_{\textup{o}}(\boldsymbol{d},\boldsymbol{\mathcal{M}})] \right]\geq \mathbb{E}_{\boldsymbol{s}}[\textrm{Conv}(r_{t,\boldsymbol{s}})].\label{eq:consider ave}
					\end{align}
					Since $r_{t,\boldsymbol{s}}$ is convex, we can change the order of the expectation and the Conv in~\eqref{eq:consider ave}. Thus we proved the converse bound in Theorem~\ref{teo}.  
					\begin{remark}
						We can also prove the converse bound in Theorem~\ref{teo} from the constraints in~\eqref{eq:take bt into xt 4},~\eqref{eq:file size} and~\eqref{eq:cache size}, by Fourier-Motzkin   elimination as it was done in~\cite{wan2016optimality,wan2016caching}.
					\end{remark}
					
					\subsection{Converse Bound for $R^*_{\textup{worst, o}}$}
					\label{sub:worse-case converse}
					We can directly extend the proof for average load in Section~\ref{sub:average converse} to the worse-case load as follows.
					\begin{align}
						R^*_{\textup{worst, o}} &= \min_{\boldsymbol{\mathcal{M}}} \max_{\boldsymbol{d}}  R^*_{\textup{o}}(\boldsymbol{d},\boldsymbol{\mathcal{M}}) \nonumber\\ &\geq \min_{\boldsymbol{\mathcal{M}}} \max_{\boldsymbol{s}\in \mathcal{S}} \mathbb{E}_{\boldsymbol{d}\in \Dc_{\boldsymbol{s}}}[ R^*_{\textup{o}}(\boldsymbol{d},\boldsymbol{\mathcal{M}})]\nonumber\\
						&\geq  \max_{\boldsymbol{s}\in \mathcal{S}} \min_{\boldsymbol{\mathcal{M}}}\mathbb{E}_{\boldsymbol{d}\in \Dc_{\boldsymbol{s}}}[ R^*_{\textup{o}}(\boldsymbol{d},\boldsymbol{\mathcal{M}})] \nonumber\\
						&\geq \max_{\boldsymbol{s}\in \mathcal{S}} \textrm{Conv}(r_{t,\boldsymbol{s}})\label{eq:convPeak}
					\end{align}
					where \eqref{eq:convPeak} follows from \eqref{eq:conv}.
					Thus we can prove the converse bound in \eqref{eq:worseImplicit}.
					
					\begin{remark}
						\label{rem:difference to shared-link}
						There are two main differences between the graphical converse bounds in~\cite{wan2016optimality,wan2016caching} for shared-link model and the ones in Theorem~\ref{teo} and Corollary~\ref{corr} for D2D model. On one hand, the D2D caching with one-shot delivery can be divided into $K$ shared-link models. The converse for the D2D model leverages the connection between these $K$ shared-link models, while in~\cite{wan2016optimality,wan2016caching} only one single shared-link model is considered. As explained in Remark~\ref{rem:improvement}, if we do not leverage this connection, we may loosen the converse bound.
						On the other hand, in the shared-link caching problem, one sub-file demanded by multiple users is treated as one sub-file. However, in the D2D caching problem, 
						the two sub-pieces  $W^{k_1,i}_{q,\Vc}$ and $W^{k_2,i}_{q,\Vc}$ where $d_{k_1}=d_{k_2}=q$ and $k_1, k_2 \notin \Vc$, which represent the sub-pieces of   sub-file $W_{q,\Vc_2}$ decoded by users $k_1$ and $k_2$ from the transmission by user $i$ respectively, are treated as two  (dependent) sub-pieces.
					\end{remark}

					\section{Proof of Lemma~\ref{lem:same coeff}}
					\label{sec:proof of same coeff}
					We assume that the set of users in $[K]\setminus\{k\}$ who have same demand as user $k$ is $\Sc$.
					Let us first focus on $W^{k,i_1}_{d_k,\Vc}$.  For a permutation of $[K]\setminus \{i_1\}$, denoted by $\mathbf{u}$, if the position of user $k$ in the vector $\mathbf{u}$ is before the position of each user in the set $\big((\Sc\cup \Vc)\setminus \{i_1,i_2\}\big )\cup \{i_2\}$ which is a subset of $([K]\setminus \{i_1\})$ with cardinality $|(\Sc\cup \Vc)\setminus \{i_1,i_2\}|+1$, $|W^{k,i_1}_{d_k,\Vc}|$ appears in the inequality in form of~\eqref{eq:converse of Xk} for this permutation. 
					
					Similarly, let us then focus on $W^{k,i_2}_{d_k,\Vc}$. For a permutation of $[K]\setminus \{i_2\}$, denoted by $\mathbf{u}^{\prime}$, if the position of user $k$ in the vector $\mathbf{u}^{\prime}$ is before the position of each user in the set $\big((\Sc\cup \Vc)\setminus \{i_1,i_2\}\big )\cup \{i_1\}$ which is a subset of $([K]\setminus \{i_2\})$ with cardinality $|(\Sc\cup\Vc)\setminus \{i_1,i_2\}|+1$, $|W^{k,i_2}_{d_k,\Vc}|$ appears in the inequality in form of~\eqref{eq:converse of Xk} for this permutation. 
					
					Hence, it is obvious to see that in~\eqref{eq:summing all Xk}, the coefficient of $|W^{k,i_1}_{d_k,\Vc}|$ is equal to the one of $|W^{k,i_2}_{d_k,\Vc}|$,
					i.e., $a^{k,i_1}_{\Vc}=a^{k,i_2}_{\Vc}$.
					
					\section{Characterization of the Worst-Case Demand}\label{sec:worstDemand}
					First recall that the binomial coefficient ${n \choose m}$ is strictly increasing in $n$.
					
					For $K \geq 2N$ if every file is demanded by at least $2$ users, every user $k$ will have $N_{\textup{e}}(\boldsymbol{d}_{\backslash\{k\}}) = N$ leading demanders, which is the maximum value possible for $N_{\textup{e}}(\boldsymbol{d}_{\backslash\{k\}})$ $\forall k \in [K]$. Hence, such a demand maximizes the load.
					
					For $K < 2N$, however, it is not possible to have $N_{\textup{e}}(\boldsymbol{d}_{\backslash\{k\}}) = N$ for all users. We call a user $k$ a \emph{unique demander} if it is the only user requesting a file. Depending on whether a user $k$ is the unique demander of a file or not, notice that $N_{\textup{e}}(\boldsymbol{d}_{\backslash\{k\}}) = N_{\textup{e}}(\boldsymbol{d}) - 1$ or $N_{\textup{e}}(\boldsymbol{d}_{\backslash\{k\}}) = N_{\textup{e}}(\boldsymbol{d})$, respectively. By the monotonicity of the binomial coefficient, a worst-case demand must have the maximum possible number of different demands, i.e., $N_{\textup{e}}(\boldsymbol{d}) = \min\{N,K\}$. Hence, $N_{\textup{e}}(\boldsymbol{d}) = N$ for $N \leq K < 2N$ and $N_{\textup{e}}(\boldsymbol{d}) = K$ for $K \leq N$ must hold. This already proves the case where $K \leq N$.

					For $N < K < 2N$, the worst-case request vector should satisfy $N_{\textup{e}}(\boldsymbol{d}) = N$ as argued above. This implies that for a user $k$ either $N_{\textup{e}}(\boldsymbol{d}_{\backslash\{k\}}) = N - 1$ or $N_{\textup{e}}(\boldsymbol{d}_{\backslash\{k\}}) = N$ should hold. Again, by the monotonicity of the binomial coefficient, a worst-case request must have the minimum number of unique demanders possible. For a worst-case request vector $\boldsymbol{d}$ with the minimum number of unique demanders which satisfies    $N_{\textup{e}}(\boldsymbol{d}) = N$, each file cannot be demanded by more than two users. Thus there are  $K-N$  files each of which is demanded by two users while each of  the remaining $2N-K$ files is demanded by exactly one user, to satisfy $\sum_{n=1}^{N}s_n(\boldsymbol{d})=K$ (i.e., there are $K$ requests). 
					Thus, we prove the case where $N < K < 2N$.
					
					Exemplary request vectors to illustrate these three cases for $N = 3$ can be $(1, 2)$, $(1, 2, 3, 1, 2)$, $(1, 2, 3, 1, 2, 3, 1)$ for $K = \{2, 5, 7\}$, respectively.
					
					\section{Order Optimality of the Proposed Scheme}\label{sec:orderOpt}
					We only show the order optimality for the average case. The same result can be proved for the worst-case by following similar steps as we present in the following.
					
					First, we notice that the load of a transmission that satisfies users' demands from a server with the whole library cannot be higher than the sum-load of transmissions from users' caches. That is to say, we have that $R^*_{\textup{ave}} \geq R^{\textup{sl*}}_{\textup{ave}}$. Furthermore, we observe that $R^{\textup{sl*}}_{\textup{ave}} \geq \frac{t}{t+1} R^*_{\textup{ave, o}}$ by the following:
					\begin{align}
						\frac{t}{t+1} R^*_{\textup{ave, o}} &= \mathbb{E}_{\boldsymbol{d}}\left[ \frac{\frac{1}{t+1}\binom{K-1}{t}-\frac{1}{K}\frac{1}{t+1}\sum_{i=1}^K\binom{K-1-N_{\textup{e}}(\boldsymbol{d}_{\backslash\{i\}})}{t}}{\frac{1}{t}\binom{K-1}{t-1}}\right]\nonumber\\
						&= \mathbb{E}_{\boldsymbol{d}}\left[ \frac{\frac{1}{K}\binom{K}{t+1}-\frac{1}{K}\sum_{i=1}^K\frac{1}{K-N_{\textup{e}}(\boldsymbol{d}_{\backslash\{i\}})}\binom{K-N_{\textup{e}}(\boldsymbol{d}_{\backslash\{i\}})}{t+1}}{\frac{1}{K}\binom{K}{t}}\right]\nonumber\\
						&\leq \mathbb{E}_{\boldsymbol{d}}\left[ \frac{\binom{K}{t+1}-\min_{i}\frac{K}{K-N_{\textup{e}}(\boldsymbol{d}_{\backslash\{i\}})}\binom{K-N_{\textup{e}}(\boldsymbol{d}_{\backslash\{i\}})}{t+1}}{\binom{K}{t}}\right]\nonumber\\	
						&\leq \mathbb{E}_{\boldsymbol{d}}\left[ \frac{\binom{K}{t+1}-\binom{K-N_{\textup{e}}(\boldsymbol{d})}{t+1}}{\binom{K}{t}}\right]\label{eq:orderoptIneq}\\
						&= R^{\textup{sl*}}_{\textup{ave}},\nonumber
					\end{align}
					where \eqref{eq:orderoptIneq} is due to $1 \leq N_{\textup{e}}(\boldsymbol{d}_{\backslash\{i\}}) \leq N_{\textup{e}}(\boldsymbol{d})$ for all $i \in [K]$.
					
					Therefore, we see that $R^*_{\textup{ave, o}} \geq R^*_{\textup{ave}} \geq \frac{t}{t+1} R^*_{\textup{ave, o}} \geq \frac{1}{2} R^*_{\textup{ave, o}}$, which concludes the proof of order optimality within a factor of $2$ under the constraint of uncoded placement. In addition, by the proved order optimality of the shared-link scheme within a factor of $2$ \cite{yu2017characterizing} for uncoded placement, it immediately follows that the achieved load is within a factor of $4$ of the information-theoretic optimum. 
					\bibliographystyle{IEEEtran}
					\bibliography{pub}

				\end{document}

%% file: header.tex
\usepackage{amsfonts, amscd, mathrsfs, amsopn, bbm, bbold, multicol, relsize}
\usepackage{mathrsfs}
\usepackage[psamsfonts]{amssymb}
\usepackage{mathbbol}
\usepackage[mathscr]{eucal}
\usepackage[cmex10]{amsmath}
\usepackage[amsmath,thmmarks]{ntheorem}
\usepackage{mathtools}
\usepackage{cite}
\usepackage{tikz}
\usepackage{graphicx}

\theoremstyle{plain}
\theoremheaderfont{\itshape\bfseries}
\theorembodyfont{\normalfont\itshape}
\theoremseparator{:}
\theoremsymbol{}
\newtheorem{theorem}{Theorem}
\newtheorem{lemma}{Lemma}
\newtheorem{corollary}{Corollary}

\newtheorem{definition}{Definition}

\newtheorem{remark}{Remark}

\theoremstyle{nonumberplain}

\theoremstyle{nonumberplain}
\theoremheaderfont{\itshape}
\theorembodyfont{\normalfont}
\theoremseparator{:}
\theoremsymbol{}

\theoremstyle{plain}
\theoremheaderfont{\itshape\bfseries}
\theorembodyfont{\normalfont}
\theoremseparator{:}

\theoremstyle{nonumberplain}
\theoremsymbol{\rule{1ex}{1ex}}

\newlength\fheight
\newlength\fwidth
\DeclarePairedDelimiter\abs{\lvert}{\rvert}

\newcommand\restr[2]{{
  \left.\kern-\nulldelimiterspace 
  #1 
  \right|_{#2} 
  }}